\documentclass[final,3p]{elsarticle}

\usepackage[T1]{fontenc}
\usepackage{amssymb}
\usepackage{amsmath}
\usepackage{upgreek}
\usepackage{bbm}
\usepackage{xfrac}

\usepackage{tikz}
\usepackage{pgfplots}
\pgfplotsset{compat=1.18}

\usepackage{caption}
\usepackage{subcaption}
\usepackage{algorithm,algpseudocode}
\usepackage{listings}
\usepackage{xcolor}
\usepackage{multirow, threeparttable}
\usepackage{adjustbox}
\usepackage{hyperref}
\usepackage{bm}

\usepackage{array}
\usepackage{diagbox}

\definecolor{codegreen}{rgb}{0,0.6,0}
\definecolor{codegray}{rgb}{0.5,0.5,0.5}
\definecolor{codepurple}{rgb}{0.58,0,0.82}
\definecolor{backcolour}{rgb}{0.95,0.95,0.92}

\lstdefinelanguage{myc++} 
{
  language=C++,
  captionpos=b,
  frame=single,
  basicstyle=\ttfamily,
  keywordstyle=\bfseries\ttfamily,
  commentstyle=\color{gray}\ttfamily,
  stringstyle=\rmfamily,
  numbers=left,
  numberstyle=\small,
  stepnumber=1,
  numbersep=8pt,
  breaklines=true,
  lineskip=1pt,
  frame=L,
  escapechar=\$
}

\newcommand{\R}{{\mathbb{R}}} % Real numbers
\newcommand{\N}{{\mathbb{N}}}
\newcommand{\controlVec}{{\boldsymbol{\alpha}}}

\newcommand{\objective}{{\mathcal{J}}}
\newcommand{\constraint}{{\mathcal{G}}}

\newcommand{\transpose}{{^\mathsf{T}}}

\newcommand{\Lagrangian}{{\mathcal{L}}}

\newcommand{\adjointsVec}{{\boldsymbol{\varphi}}}

\begin{document}

\begin{frontmatter}

\title{Mean-Flow Adjoint Sensitivity Analysis of Unsteady Flow Around Porous Cylinders Using a Homogenized Lattice Boltzmann Method}

\address[mvm]{Institute for Mechanical Process Engineering and Mechanics, Karlsruhe Institute of Technology, Karlsruhe, Germany}
\address[lbrg]{Lattice Boltzmann Research Group, Karlsruhe Institute of Technology, Karlsruhe, Germany}
\address[ianm]{Institute of Applied and Numerical Mathematics, Karlsruhe Institute of Technology, Karlsruhe, Germany}

\author[mvm,lbrg]{Shota Ito}
\author[mvm,lbrg]{Johannes L. Grafen}
\author[mvm,lbrg]{Fedor Bukreev}
\author[lbrg,ianm]{Adrian Kummerländer}
\author[mvm,lbrg,ianm]{Mathias J. Krause}

\begin{abstract}
Adjoint-based sensitivity analysis is an indispensable tool for large-scale fluid-dynamic design and distributed control problems, yet its application to unsteady and turbulent flows is frequently hindered by the prohibitive memory footprint of transient checkpointing and the divergence of gradients in chaotic regimes.
To address these computational bottlenecks, this paper presents a mean-flow adjoint sensitivity analysis framework for unsteady flows around porous cylinders using the homogenized lattice Boltzmann method (HLBM).
Within this framework, solid structures are efficiently modeled as local porous media utilizing a Brinkman penalization approach.
We systematically investigate HLBM-based adjoint gradients for drag and energy dissipation objective functionals, transitioning from steady laminar to unsteady, and finally to turbulent flow regimes.
For the turbulent case at \textit{Re} = 3900, a proof-of-concept is
conducted where the framework relies on automatic differentiation to automatically generate adjoint kernels containing subgrid-scale (SGS) turbulence models for large eddy simulations (LES), circumventing manual derivation and allowing for a direct comparison against the frozen turbulence assumption (FTA).
\end{abstract}

\end{frontmatter}

\section{Introduction}

The optimization of fluid dynamic design and the optimal control of flow problems pose key disciplines in aerodynamic engineering~\cite{cheylan_shape_2019, jalali_khouzani_airfoil_2023, stuck2012adjoint}. 
Therein, it is an established procedure to couple methods of mathematical optimization and computational fluid dynamics (CFD) simulations to perform design improvements in an automated manner, while enforcing physicality of the optimized solution by numerically approximating the governing partial differential equations of the underlying flow phenomenon~\cite{dugast_topology_2018, krause_adjoint-based_2013}.
Therein, gradient-based optimization (GBO) is of particular interest as the iterative execution of CFD simulations for industrial-scale problems dominates the computational cost, where GBO, utilizing local gradient information, in general outperforms other approaches in terms of convergence rates~\cite{laniewski-wollk_adjoint_2016, ito_identification_2024}.

Adjoint-based sensitivity analysis has established itself as an indispensable tool for solving large-scale gradient-based fluid-dynamic design and flow control problems~\cite{chen_local--time_2017, ito_geometry_2026}.
Because the computational cost of evaluating an objective gradient via the adjoint method remains independent of the number of design variables, it is suited for distributed control problems such as fluid topology optimization~\cite{borrvall2003, pingen_adjoint_2009, yaji_topology_2016, liu_discrete_2014} or shape optimization~\cite{mohammadi2004shape, YAJI2014158}.
Within this context, the \textit{lattice Boltzmann method} (LBM) represents a highly attractive discretization scheme due to its great suitability for parallel computing~\cite{krause_openlbopen_2021, kummerlander_efficient_vocal_2026, kummerlander_efficient_2026-1}.
Beyond offering remarkable computational efficiency within iterative optimization loops, LBM is highly compatible with the adjoint analysis, as the corresponding adjoint problem can be again discretized by an LBM, providing the same performance benefits~\cite{ito_generation_2026}.

To incorporate the solid body in the fluid topology optimization, a \textit{Brinkman penalization approach} is often applied, where the pressure drop is modeled by a local porous medium imitating the immersed solid structure~\cite{ito_geometry_2026, yaji_topology_2016, klemens_solving_2020}.
By controlling the permeability of the porous medium, the topology can be optimized regarding different objective functionals such as drag or pressure drop.
Krause \textit{et al.}~\cite{KRAUSE20171} proposed the \textit{homogenized lattice Boltzmann method} (HLBM), extended by Kummerländer \textit{et al.}~\cite{kummerlander_efficient_2026-1}, which efficiently employs the Brinkman penalization to model solid structures in the fluid.
In prior works, the HLBM method has been successfully applied for topology optimization problems, such as in~\cite{ito_geometry_2026, klemens_solving_2020}.

The application of adjoint sensitivity analysis to transient flow control, especially in LBM, remains very limited in the existing literature.
In the realm of unsteady flows, standard transient adjoint simulations require checkpointing or storing the entire forward state history to be traversed in reverse time, creating an often prohibitive memory bottleneck~\cite{NORGAARD2016291, meliga2014}. 
Prior work reports efficient checkpointing approaches such as the \textit{local-in-time method}, in which checkpointing is combined with approximations of unsteady adjoint solutions to reduce the memory footprint in the adjoint simulation~\cite{chen_local--time_2017,YAMALEEV20105394} .
However, for large-scale simulations, especially for three-dimensional problems, the storage of the primal state solution becomes quickly computationally unfeasible~\cite{yaji2018large}.
Especially for turbulent flows, computational resources become prohibitively large when resolving transient structures.
Furthermore, the chaotic flow dynamics lead to diverging solution trajectories and thus to failure of conventional gradient computation, requiring more sophisticated and numerically expensive approaches, such as the \textit{least squares shadowing} method~\cite{meliga2014, LSS2018}.

In classical approaches, the adjoint analysis is applied to mean flow solutions of turbulent flows by combining it with \textit{Reynolds averaged Navier-Stokes} (RANS) simulations to circumvent the diverging gradients~\cite{MARTA2013102, papoutsis2016continuous, computation6010005}.
Additionally, by driving the reversed-time adjoint equations using the time-averaged primal flow field rather than instantaneous transient snapshots, Dwight~\textit{et al.} demonstrated that computationally tractable and physically meaningful surface sensitivities could be recovered~\cite{dwight2006}. 
Meliga \textit{et al.} investigated the adjoint 2D squared cylinder flow using a finite element method for Reynolds numbers up to $22000$.
A RANS model is applied in the primal simulation, where the adjoint problem is constructed on the averaged solutions to handle the chaotic flow behavior in the turbulent case~\cite{meliga2014}.
In LBM, Cheylan \textit{et al.} proposed the mean-flow adjoint-based sensitivity analysis for aerodynamic shape optimization to address drag minimization problems for a vehicle~\cite{cheylan_shape_2019}.
Therein, the primal state solution is averaged and used to evaluate the Jacobian expressions in the adjoint simulation, in which the controlled body shape is incorporated into the flow field via adjoint curvature-capturing boundary conditions.

To the best of the authors' knowledge, the investigation of mean-flow-based adjoint sensitivities for LBM simulations is not yet applied in the context of topology optimization, e.g., for a Brinkman penalization model such as HLBM.
Therefore, the present work systematically investigates HLBM-based adjoint gradients for a porous cylinder with drag-type functionals, from steady laminar to unsteady and finally turbulent flow regimes.
Both works~\cite{cheylan_shape_2019} and~\cite{meliga2014}, neglect the subgrid-scale (SGS) turbulence model in the adjoint sensitivity analysis by employing the \textit{frozen turbulence assumption} (FTA) due to the challenging manual derivation of the non-linear adjoint \textit{large eddy simulation} (LES) based on the Smagorinsky model, despite the loss in gradient accuracy~\cite{cheylan_shape_2019,computation6010005}.
So far, the comparison of computed gradients from fully differentiated turbulence models and those obtained by the FTA has been reported only for RANS models~\cite{MARTA2013102, computation6010005}.
This manuscript demonstrates preliminary comparison results of adjoint Smagorinsky-LES models and the FTA for the turbulent flow past a 3D cylinder.
The framework developed by Ito \textit{et al.} is used, where the adjoint kernels, containing SGS turbulence models, are automatically generated from their primal implementations using \textit{automatic differentiation} (AD), omitting the need for manual derivation~\cite{ito_generation_2026}.
The research questions of the present work are formulated as follows:

\begin{enumerate}
    \item Can HLBM adjoints compute accurate drag-type sensitivities for a porous cylinder flow?
    \item How do mean-flow adjoint gradients behave for unsteady vortex shedding using the HLBM setting?
    \item Can the same framework be extended to 3D and to LES-level turbulent flow as a proof of concept? And, how does the FTA alter the adjoint sensitivities for the present LES simulation?
\end{enumerate}
Consequently, the primary novelties of this paper consists of the first sensitivity evaluation of an HLBM-adjoint for porous cylinder flow, the first application of HLBM mean-flow adjoints to unsteady flows across multiple distinct objective functionals, and the demonstration of a 3D turbulent adjoint cylinder flow at a Reynolds number of $3900$.

The remainder of this paper is structured as follows: Sec.~\ref{sec:lbm} introduces the governing primal physical models for laminar and turbulent regimes. 
Sec.~\ref{sec:adjoint} details the steady and unsteady mean-flow adjoint sensitivity frameworks. 
Sec.~\ref{sec:objective} defines the investigated drag and dissipation objective functionals. 
Finally, Sec.~\ref{sec:results} presents the results containing numerical validation and evaluation studies, followed by concluding remarks in Sec.~\ref{sec:conclusion}.
\section{Primal modeling\label{sec:lbm}}

This section provides the used primal physical models first for the laminar flow in Sec.~\ref{sec:laminar} and then for the turbulent flow in Sec.~\ref{sec:turbulent}.

\subsection{Laminar model}\label{sec:laminar}

For the laminar case, the incompressible Navier-Stokes equations describe the motion of fluid.
To incorporate the solid cylinder, the Brinkman-type Navier-Stokes equation~\cite{simonis2025homogenized} is used

\begin{align}
\begin{cases}
\bm{\nabla} \cdot \bm{u} =0, & \quad \text{in } \Omega \times I, \\
\frac{\partial \bm{u}}{\partial t} + \bm{u} \cdot \bm{\nabla} \bm{u} = -\frac{\bm{\nabla} p}{\rho} + \nu \bm{\nabla}^2 \bm{u} - \frac{\nu}{K} \bm{u}, & \quad \text{in } \Omega \times I, \label{eq:btnse}
\end{cases}
\end{align}

\noindent where $\bm{u}$ is the velocity, $t$ the time, $p$ the pressure, $\rho$ the density, $\nu$ the molecular viscosity, and $K$ the permeability modeling the porous medium submerged in the fluid.
The above equation holds in the $d$-dimensional spatial domain $\Omega \subseteq \R^d$ for the time duration $I \subseteq R_{>0}$.
The HLBM~\cite{KRAUSE20171} with the Bhatnagar-Groos-Krook (BGK)~\cite{Bhatnagar.1954} collision model approximate \eqref{eq:btnse} in the hydrodynamic limit, given as

\begin{align}
    f_i(\bm{x}+\bm{\xi}_i\triangle t, t + \triangle t) = f_i(\bm{x}, t) - \frac{\triangle t}{\tau}(f_i(\bm{x}, t) - f_i^{\mathrm{eq}}(\bm{x}, t)), \quad \text{in } \Omega_{\triangle x} \times I_{\triangle t}, \label{eq:hlbm}
\end{align}

\noindent where $f_i \in \R$ is the particle distribution function transported along $q \in \N_{>0}$ discrete velocity directions with the particle velocity $\bm{\xi}_i \in \R^d$ corresponding to the chosen DdQq discretization model~\cite{Qian.1992}.
The D2Q9 stencil and the D3Q19 stencil have been used for the numerical experiments in 2D and 3D, respectively.
Therein, the particle velocity is given such that on a regular lattice $\Omega_{\triangle x} \subseteq \Omega$ with the grid spacing $\triangle x \in \R_{>0}$, the particle distribution function is advected exactly to its neighbor position after a single time step $\triangle t \in I_{\triangle t} \subseteq I$. 
The equilibrium distribution is given as

\begin{align}
    f_i^{\mathrm{eq}}(\bm{x},t) = \omega_i  \rho \left(1+\frac{\bm{\xi}_i\cdot \Tilde{\bm{u}}}{c_s^2}+\frac{(\Tilde{\bm{u}}\cdot\bm{\xi}_i)^2}{2c_s^4}-\frac{\Tilde{\bm{u}}\cdot \Tilde{\bm{u}}}{2c_s^2}\right),
\end{align}

\noindent where $\omega_i \in \R$ are corresponding weights of the velocity discretization model, $\Tilde{\bm{u}} = d\cdot\bm{u}$ is the modified velocity with $d = 1 - \frac{\triangle x^2 \tau \nu}{K}$ being the lattice porosity linked to the permeability~\cite{klemens_solving_2020, spaid1997lattice}, and $c_s = \frac{1}{\sqrt{3}}\frac{\triangle t}{\triangle x}$ is the lattice speed of sound~\cite{Kruger.2017}.
The statistic moments connect the particle distribution functions to the fluid density and velocity via

\begin{align}
    \rho =& \sum_{i=0}^{q-1} f_i, \\
    \bm{u} =& \frac{1}{\rho} \sum_{i=0}^{q-1} \bm{\xi}_i f_i.
\end{align}

\noindent Finally, the fluid viscosity is recovered via

\begin{align}
    \nu = c_s^2 \left(\tau - \frac{\triangle t}{2}\right).
\end{align}

\subsection{Turbulent model}\label{sec:turbulent}

In the turbulent case, a SGS model for LES is added to the incompressible Brinkman-type Navier-Stokes equations, given as

\begin{align}
\begin{cases}
\bm{\nabla} \cdot \bm{u} =0, & \quad \text{in } \Omega \times I, \\
\frac{\partial \bm{u}}{\partial t} + \bm{u} \cdot \bm{\nabla} \bm{u} = -\frac{\bm{\nabla} p}{\rho} + \nu_{\mathrm{mo}} \bm{\nabla}^2 \bm{u} - \frac{\nu_{\mathrm{mo}}}{K} \bm{u} - \bm{\nabla} \cdot \mathbf{T}_{\mathrm{sgs}}, & \quad \text{in } \Omega \times I, \label{eq:btnseles}
\end{cases}
\end{align}

\noindent where the term $\bm{\nabla} \cdot \mathbf{T}_{\mathrm{sgs}}$ models the subgrid scale turbulence using the Smagorinsky-LES model~\cite{smagorinsky1963general} via

\begin{align}
    \mathbf{T}_{\mathrm{sgs}} & = -2 \nu_\mathrm{turb} \mathbf{S}, \label{eq:sgsStress}\\
    \nu_\mathrm{turb} & = \left(C_{\mathrm{S}} \triangle x \right)^2 \left|\mathbf{S}\right|, \label{eq:turbVisc}
\end{align}  
where $C_{\mathrm{S}} \in \R_{>0}$ is the Smagorinsky constant, $\triangle x$ is the filter width, and $\mathbf{S}$ is the strain rate tensor: 
\begin{equation}
    S_{\alpha\beta} = \frac{1}{2} \left( \frac{\partial u_{\alpha}}{\partial x_\beta} + \frac{\partial u_{\beta}}{\partial x_\alpha} \right),
\end{equation}

\noindent with $\alpha, \beta \in {1,2,3}$ being the spatial indices.
To discretize \eqref{eq:btnseles}, the HLBM method is extended by the \textit{homogenized regularized recursive lattice Boltzmann method with Smagorinsky LES model} (HRRLBM-LES) proposed by Kummerländer \textit{et al.}~\cite{kummerlander_efficient_vocal_2026}, which is based on a third-order recursive regularized collision model~\cite{coreixas2017recursive, jacob2018new}, given as

\begin{equation}
    f_{i} (\bm{x}+\bm{\xi}_i \triangle t, t+\triangle t) 
    = 
    f_{i}^{\mathrm{eq}} (\bm{x}, t) + \left( 1 - \frac{1}{\tau_{\mathrm{eff}}(\bm{x},t)} \right)f_{i}^{(1)}(\bm{x}, t), \quad \text{in } \Omega_{\triangle x} \times I_{\triangle t}, \label{eq:HRRLBM}
\end{equation}

\noindent where $\tilde{f}_{i}^{(1)}$ is the non-equilibrium distribution expanded in terms of Hermite polynomials $\mathbf{H}_i^{(n)}$ of the discrete velocity $\boldsymbol{\xi}_i$, given as

\begin{align}
    f_{i}^{(1)}(\bm{x},t) = \omega_i \sum_{n=0}^{N=3} \frac{1}{c_{\mathrm{s}}^{2n} n!} \mathbf{H}_i^{(n)} : \mathbf{a}_1^{(n)}(\bm{x},t),
\end{align}

\noindent where the Hermite expansion coefficients are defined as

\begin{align}
    \mathbf{a}_1^{(n)}(\bm{x},t)=\sum_{i=0}^{q-1}\mathbf{H}_i^{(n)}f_i^{(1)}(\bm{x},t).
\end{align}

\noindent The equilibrium distribution is given as

\begin{align}
    f_i^{\mathrm{eq}}(\bm{x},t) = \omega_i \left(\rho + \frac{\bm{\xi}_i \cdot \rho\, \Tilde{\bm{u}}}{c_{\mathrm{s}}^2}
    + \frac{\mathbf{H}_i^{(2)} : \widehat{\mathbf{a}}_0^{(2)}}{2 c_{\mathrm{s}}^4}
    + \frac{\mathbf{H}_i^{(3)} : \widehat{\mathbf{a}}_0^{(3)}}{6 c_{\mathrm{s}}^6}
  \right)
\end{align}

\noindent using Hermite coefficients \(\widehat{\mathbf{a}}_{0}^{(0)} = \rho(\bm{x},t)\) and \(\widehat{\mathbf{a}}_{0}^{(n)} = \widehat{\mathbf{a}}_{0}^{(n-1)} \Tilde{\bm{u}}(\bm{x},t)\).
Finally, the SGS turbulence is accounted for by locally computing the effective relaxation time $\tau_\mathrm{eff}(\bm{x},t)$ using the Smagorinsky BGK model~\cite{Malaspinas_Sagaut_2012}
\begin{equation}\label{eq:tauEff}
    \tau_\mathrm{eff}(\bm{x},t) = \frac{\nu_\mathrm{mo}+\nu_\mathrm{turb}(\bm{x},t)}{c_{\mathrm{s}}^2}\frac{\triangle t}{\triangle x^2} + \frac{1}{2}.
\end{equation}
\section{Adjoint sensitivity analysis\label{sec:adjoint}}

In this section, the optimization problem is formulated in Sec.~\ref{sec:optimization} for which the adjoint problem delivers the sensitivities.
The first-discretize-then-differentiate approach is employed in this work~\cite{gunzburger2002}.
For clarity, we separate the explanation for the adjoint-based sensitivity computation for the steady case and unsteady case in Sec.~\ref{sec:steady_opti} and Sec.~\ref{sec:unsteady_opti}, respectively.
Note that in the present work, we do not solve optimization problems numerically using optimization algorithms. 
However, we do compute objective gradients with an adjoint-based approach, for which the formulation of an optimization problem is necessary.

\subsection{Optimization problem formulation}\label{sec:optimization}
The optimization problem is formulated as

\begin{equation}
    \begin{array}{l@{\quad}rrr}
        \min\limits_{\boldsymbol{\controlVec}} \objective(\controlVec,f) & \text{while} & \constraint(\controlVec,f) = 0,
    \end{array}\label{eq:optimizationProblem}
\end{equation}

\noindent where $\bm{\alpha} \in \Omega_{\bm{\alpha}}$ contains the time-invariant permeability distribution including the porous cylinder in the design domain $\Omega_{\bm{\alpha}} \subseteq \Omega_{\triangle x}$, $f$ is the flow state, $\objective$ the objective functional, and $\constraint$ the optimization constraints being either the residual of~\eqref{eq:hlbm} or~\eqref{eq:HRRLBM}, depending on the used primal model.
The objective gradients regarding the controls are given as

\begin{align}
    \frac{\mathrm{d} \objective}{\mathrm{d} \controlVec} = \frac{\partial \objective}{\partial \controlVec} + \frac{\partial \objective}{\partial f} \frac{\partial f}{\partial \controlVec},
\end{align}

\noindent where the explicit computation of $\partial f / \partial \controlVec$ is cumbersome, especially when the number of control variables is high, as in topology optimization.
To avoid this, the Lagrangian function is used here as an augmented objective functional instead, given as

\begin{align}
    \Lagrangian = \objective + \adjointsVec\transpose \constraint,
    \label{eq:lagrangian_general}
\end{align}

\noindent where $\adjointsVec$ is the Lagrangian multiplier.
They are obtained by solving $\partial \Lagrangian / \partial f = \mathbf{0}$, which is a linear system solved independently of the number of control variables~\cite{cheylan_shape_2019}.
Finally, the total derivative of the objective regarding the controls can then be computed via

\begin{align}
    \frac{\mathrm{d}\Lagrangian}{\mathrm{d}\controlVec} =
    \frac{\mathrm{d}\objective}{\mathrm{d}\controlVec} = \frac{\partial \objective}{\partial \controlVec} + \adjointsVec\transpose \frac{\partial \constraint}{\partial \controlVec}.
\end{align}

\subsection{Steady case}\label{sec:steady_opti}
In the steady case, the discrete objective functional is evaluated for the converged primal steady state solution $f^*(\cdot)$ via $\objective = J(\bm{\alpha}, f^*(\cdot))$.
The adjoint LBM model is given as

\begin{align}\label{eq:adjoint_steady}
    \varphi_i(\bm{x}-\bm{\xi}_i \triangle t, t - \triangle t) = \sum_{j=0}^{q-1}\left(\left.\frac{\partial \mathcal{C}_j(\bm{x}, t)}{\partial f_i(\bm{x}, t)}\right|_{f_i = f_i^*(\bm{x})}\right)^\transpose \varphi_j(\bm{x}, t) - \left.\frac{\partial J}{\partial f_i(\bm{x}, t)}\right|_{f_i = f_i^*(\bm{x})},
\end{align}

\noindent where $\mathcal{C}_i$ is the substitution of the local evolution operator on the right-hand side of~\eqref{eq:hlbm} or \eqref{eq:HRRLBM}.
Note that the adjoint problem traverses in reverse time and would require the primal state variables in reverse time order to evaluate the Jacobian matrices.
In \eqref{eq:adjoint_steady}, we only use the primal state of the converged state following the work of Krause \textit{et al.}~\cite{krause_adjoint-based_2013}, who proposed this valid simplification for steady state problems.
The objective sensitivities are computed as

\begin{align}\label{eq:gradients_steady}
     \frac{\mathrm{d} \objective}{\mathrm{d} \controlVec} = \left.\frac{\partial J}{\partial \bm{\alpha}}\right|_{f_i = f_i^*(\bm{x})} - \sum_{\bm{x} \in \Omega_{\bm{\alpha}}}\sum_{i=0}^{q-1}\left(\left.\frac{\partial \mathcal{C}_i(\bm{x},t)}{\partial \bm{\alpha}}\right|_{f_i = f_i^*(\bm{x})}\right)^\transpose \varphi_i^*(\bm{x}),
\end{align}

\noindent where $\varphi_i^*(\bm{x})$ is the converged mean-flow adjoint steady state solution.
To obtain the explicit formulation of the Jacobian matrices, the framework proposed by Ito \textit{et al.}~\cite{ito_generation_2026} using automatic differentiation (AD) is used to omit the tedious derivations.
For more detailed explanations, the readers are referred to that work to keep the methodology section in the present paper concise.

\subsection{Unsteady case}\label{sec:unsteady_opti}

In the unsteady case, the objective functional is defined as a tracking average-type functional:

\begin{align}
    \objective = \overline{J(\bm{\alpha}, f(\cdot, t))} = \frac{1}{t_e - t_s}\sum_{t=t_s}^{t_e} J(\bm{\alpha}, f(\cdot, t)),
    \label{eq:averageObjective}
\end{align}

\noindent where $t_s, t_e \in I_{\triangle t}$ are the start and end time steps of the averaging duration, respectively.
In the present work, we employ the approach proposed by Cheylan \textit{et al.}~\cite{cheylan_shape_2019}, namely the \textit{mean-flow adjoint LBM}.
That is, in the unsteady adjoint problem, one would need to store the entire primal solution history to access it in the reverse order during the adjoint simulation to evaluate the Jacobian matrices in~\eqref{eq:adjoint_steady}.
Instead, in the mean-flow adjoint approach, we compute the tracking average of the primal populations $\overline{f(\cdot, t)}$ as in~\eqref{eq:averageObjective} and evaluate the Jacobian matrices for the primal mean-flow solution, i.e.:

\begin{align}\label{eq:adjoint_unsteady}
    \varphi_i(\bm{x}-\bm{\xi}_i \triangle t, t - \triangle t) = \sum_{j=0}^{q-1}\left(\left.\frac{\partial \mathcal{C}_j(\bm{x}, t)}{\partial f_i(\bm{x}, t)}\right|_{f_i = \overline{f_i(\bm{x}, t)}}\right)^\transpose \varphi_j(\bm{x}, t) - \overline{\left.\frac{\partial J}{\partial f_i(\bm{x}, t)}\right|_{f_i=f_i(\bm{x},t)}},
\end{align}

\noindent where the adjoint source term gives the average partial derivative of the objective regarding the particle distribution functions, again computed as in~\eqref{eq:averageObjective}.
The objective sensitivities are computed as

\begin{align}\label{eq:gradients_unsteady}
     \frac{\mathrm{d} \objective}{\mathrm{d} \controlVec} = \overline{\left.\frac{\partial J}{\partial \bm{\alpha}}\right|_{f_i=f_i(\bm{x},t)}} - \sum_{\bm{x} \in \Omega_{\bm{\alpha}}}\sum_{i=0}^{q-1}\left(\left.\frac{\partial \mathcal{C}_i(\bm{x},t)}{\partial \bm{\alpha}}\right|_{f_i = \overline{f_i(\bm{x}, t)}}\right)^\transpose \varphi_i^*(\bm{x}).
\end{align}

\noindent Note, that if the objective functional is linear regarding $f$, then the following equality of $\overline{J(\bm{\alpha}, f)} = J(\bm{\alpha}, \overline{f})$ holds, which omits the need for evaluating the objective every $t \in [t_s, t_e]$~\cite{meliga2014}.
Similarly, if $J$ and $\mathcal{C}$ are linear regarding $f$, then the same equality holds for the terms $\partial J/\partial f$, $\partial J/\partial \bm{\alpha}$, $\partial \mathcal{C}/\partial f$, and $\partial \mathcal{C}/\partial \bm{\alpha}$.
If $J$ or $\mathcal{C}$ are quadratic regarding $f$, then the identity is only valid for the partial derivatives regarding the state.
All objective functionals and collision operators which have a higher non-linearity than the second order would need to track the average of the above listed terms every time step during the averaging window.
The full explanation for this is given in~\ref{app:tempAvg}.
For both collision operators given in the previous section, the temporal averaging of their partial derivatives required in ~\eqref{eq:adjoint_unsteady} and~\eqref{eq:gradients_unsteady} would be necessary.
However, this would necessitate the storage of full Jacobian matrices of the collision operators on every grid position in the entire simulation domain, leading to a critical increase of the memory bandwidth during the primal and adjoint simulation.
Therefore, we propose to track the averaged entities for the scalar objective functional terms only as given in the algorithms above.

\subsection{Algorithm}

This section briefly summarizes the entire workflow to compute the adjoint-based objective sensitivities regarding the control variables for the steady and unsteady cases.
The algorithm in~\ref{alg:opti} illustrates the process employed in the current work.
The workflow is fundamentally split into three phases: the primal evaluation, the adjoint evaluation, and the final gradient assembly. First, the primal LBM is solved forward in time. For steady problems, the simulation simply runs until the local state updates fall below a prescribed convergence tolerance $\varepsilon$. For unsteady or turbulent configurations, the simulation undergoes an initial transient spin-up phase ($t\leq t_s$). 
Afterward, a time-averaging process is initiated, and the primal simulation terminates only when the statistical fluctuations of these averaged quantities satisfy a user-defined convergence criterion.
Once the primal flow field is obtained, the adjoint simulation is executed.
In the steady case, the adjoint equations are evaluated using the converged primal steady state.
For the unsteady case, to circumvent the massive memory footprint associated with exact transient checkpointing, the algorithm employs a mean-flow-based adjoint approach as described in Sec.~\ref{sec:unsteady_opti}.
Here, the adjoint equations are driven by the time-averaged primal flow field and solved until the adjoint variables themselves reach a steady state. 
Finally, the objective sensitivities are assembled by coupling the control-dependent Jacobians with the corresponding converged primal and adjoint states.

\begin{algorithm}[t!]
\caption{Adjoint-based sensitivity computation}\label{alg:opti}
\begin{algorithmic}[1]
\Procedure{Compute }{$\mathrm{d}\objective/\mathrm{d}\controlVec$}
\State Choose a convergence tolerance $\varepsilon$
\State Solve primal problem $\constraint(\bm{\alpha}, f)$ \Comment{\eqref{eq:hlbm} or \eqref{eq:HRRLBM}}
\If{Steady}
\State Terminate simulation when steady state solution is reached \Comment{state update $< \varepsilon$}
\State Solve adjoint problem \Comment{\eqref{eq:adjoint_steady}}
\Else
\State Start computing tracking average after $t > t_s$\Comment{E.g.:~\eqref{eq:averageObjective}}
\State Terminate simulation when fluctuation of averaged entities is below $\varepsilon$
\State Solve mean-flow based adjoint problem \Comment{\eqref{eq:adjoint_unsteady}}
\EndIf
\State Terminate adjoint simulation when steady state solution is reached \Comment{state update $< \varepsilon$}
\If{Steady}
\State Compute objective sensitivities using primal and adjoint results \Comment{\eqref{eq:gradients_steady}} 
\Else
\State Compute objective sensitivities using primal and mean-flow adjoint results \Comment{\eqref{eq:gradients_unsteady}} 
\EndIf
\EndProcedure
\end{algorithmic}
\label{alg:opti}
\end{algorithm}

All described methods and conducted numerical experiments are realized within the open-source C++ library OpenLB~\cite{krause_openlbopen_2021}.
\section{Objective functionals\label{sec:objective}}

This section specifies the three investigated objective functionals for the cylinder flow. Depending on the flow regime, these instantaneous objective functionals $J(t)$ are either evaluated at the converged steady state or time-averaged over a specific duration, as outlined in Sec.~\ref{sec:adjoint}.

\subsection{Drag (Momentum exchange algorithm)}
The first functional evaluates the drag force exerted on the cylinder using the momentum exchange algorithm (MEA)~\cite{laddNumericalSimulationsParticulate1994, laddNumericalSimulationsParticulate1994a}. The algorithm calculates the force based on the momentum variation of the particle distribution functions hitting the solid boundaries. The instantaneous objective functional is defined as

\begin{align}
    J_{\mathrm{MEA}}(t) = \left[\frac{1}{A_{\Gamma_{\triangle x, \mathrm{s}}}} \sum_{\bm{x} \in \Gamma_{\triangle x, \mathrm{s}}} \sum_{i=0}^{q-1} \bm{\xi}_i \left( f_i(\bm{x} + \bm{\xi}_i\triangle t, t) - f_{\bar{i}}(\bm{x}, t) \right)\right] \cdot \bm{n}, \quad \text{in }  \Gamma_{\triangle x, \mathrm{s}}, \label{eq:obj_mea}
\end{align}

\noindent where $A_{\Gamma_{\triangle x, \mathrm{s}}}$ represents the surface area (or the perimeter in 2D) of the cylinder, $\Gamma_{\triangle x, \mathrm{s}} \subseteq \Omega_{\triangle x}$ denotes the boundary nodes associated with the cylinder surface, $\bar{i}$ indicates the discrete velocity direction opposite to $i$ (i.e., $\bm{\xi}_{\bar{i}} = -\bm{\xi}_i$), and $\bm{n}$ gives the normal vector to get the drag force component in the main flow direction.

\subsection{Drag (Far field approximation)}
Alternatively, the drag can be computed using a far-field approximation (FFA) based on a control volume approach as proposed by Onorato \textit{et al.}~\cite{ONORATO1986317}. Instead of evaluating the forces directly at the obstacle surface, this functional calculates the momentum deficit and pressure drop across a boundary $\Gamma_{\triangle x, J} \subseteq \Omega_{\triangle x, J}$ enclosing the objective domain $\Omega_{\triangle x, J} \subseteq \Omega_{\triangle x}$:

\begin{align}
    J_{\mathrm{FFA}}(t) = \sum_{\bm{x} \in \Gamma_{\triangle x, J}} \rho(\bm{x},t) \left[ u_x(\bm{x},t) u_0 - u_x(\bm{x},t)^2 \right] + \sum_{\bm{x} \in \Gamma_{\triangle x, J}} (p_0 - p(\bm{x},t)), \quad \text{in }  \Gamma_{\triangle x, J}, \label{eq:obj_ffa}
\end{align}

\noindent where $u_x$ is the macroscopic velocity component in the primary flow direction, $u_0$ is the reference free-stream velocity, and $p_0$ is the reference far-field pressure.

\subsection{Dissipation (Porous and Viscous contribution)}
The third functional evaluates the total energy dissipation within the objective domain $\Omega_{\triangle x, J} \subseteq \Omega_{\triangle x}$, following the formulation by Wang \textit{et al.}~\cite{Wang2011}. This functional accounts for both the viscous dissipation of the fluid and the porous dissipation induced by the Brinkman penalization term:

\begin{align}
    J_{\mathrm{DISS}}(t) = \sum_{\bm{x} \in \Omega_J} \left[ 2\nu (\mathbf{S}(\bm{x},t) : \mathbf{S}(\bm{x},t)) + \frac{\nu}{K} \bm{u}(\bm{x},t) \cdot \bm{u}(\bm{x},t) \right], \quad \text{in }  \Omega_{\triangle x, J},\label{eq:obj_diss}
\end{align}

\noindent where $\mathbf{S}$ is the macroscopic strain rate tensor. In the LBM framework, the strain rate components $S_{\alpha\beta}$ can be locally recovered from the non-equilibrium part of the particle distribution functions, defined as $f_i^{\mathrm{neq}} = f_i - f_i^{\mathrm{eq}}$, without the need for finite-difference approximations:

\begin{align}
    S_{\alpha\beta}(\bm{x},t) = -\frac{1}{2\tau c_s^2} \sum_{i=0}^{q-1} \xi_{i\alpha} \xi_{i\beta} f_i^{\mathrm{neq}}(\bm{x},t).
\end{align}
\section{Results\label{sec:results}}

The 2D and 3D flow around a cylinder is investigated, where the cylinder is modeled via the Brinkman-penalization.
First, the validity of the HLBM is demonstrated by comparing the results with no-slip boundary conditions applied on the cylinder surface for the 2D Schäfer-Turek benchmark case~\cite{schafer1996benchmark} in Sec.~\ref{sec:primalValidation}.
Sec.~\ref{sec:adjointValidation} present the validation results of the adjoint sensitivities for the steady laminar case at $Re = 20$ in 2D and $Re = 40$ in 3D with the Reynolds number defined as $Re = \frac{u_{x,\infty} D}{\nu}$, where $u_{x,\infty}$ is the inflow velocity and $D$ the diameter of the cylinder.
Then, Sec.~\ref{sec:meanFlowAdjoints} contains the results for the laminar but unsteady case at $Re = 100$, where the gradients are computed by the mean-flow adjoint approach.
Finally, Sec.~\ref{sec:proof-of-concept} demonstrates preliminary results for the unsteady turbulent case at $Re = 3900$, where the adjoint HRRLBM-LES model is compared with the FTA~\cite{MARTA2013102}.

\subsection{Validation of the primal model: Drag of a 2D porous cylinder flow}\label{sec:primalValidation}

To demonstrate the feasibility of the primal problem with the HLBM approach for drag-type objective functionals, the 2D Schäfer-Turek benchmark case~\cite{schafer1996benchmark} is studied.
The simulation parameters and domain sizes are chosen identically as defined in the benchmark case. 
The cylinder diameter $D$ is resolved by $N = 20$ cells, and the lattice porosity in the cylinder is set to $d = 0$ and $d = 1$ in the fluid bulk.
The results are compared regarding the drag coefficient, defined as

\begin{align}
    c_D = \frac{2 F_x}{\rho u_{x,\infty}^2 D}.
\end{align}

\noindent Additionally to the benchmark reference, no-slip boundary conditions on the cylinder surface realized by the bounce back method and interpolated bounce back approach~\cite{bouzidi2001momentum} are added in the comparison study.

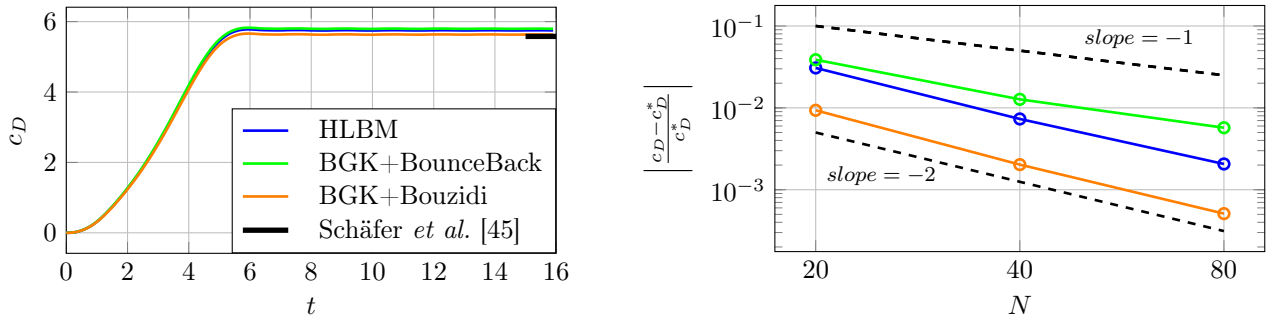
\begin{figure}[h!]
  \begin{center}
    \begin{subfigure}[t]{.49\textwidth}
    \begin{tikzpicture}
      \begin{axis}[
          width=\linewidth*1, % Scale the plot to \linewidth
          height=\linewidth*0.6,
          %xmode=log,
          %ymode=log,
          xlabel=$t$, % Set the labels
          ylabel=$c_D$,
          xlabel style={scale=1},
          ylabel style={scale=1},
          %x unit=\si{\degree}, % S   return util::pow(output, 1. / P);et the respective units
          %y unit=\si{\pascal},
          legend cell align=left,
          legend style={column sep=0.3cm,nodes={transform shape},at={(1,0)},anchor=south east}, % Put the legend below the plot
          x tick label style={anchor=north}, % Display labels sideways
          y tick label style={},
          xmin=0,
          xmax=16,
          grid
        ]
        \addplot [line width=1pt,solid,color=blue] table[x index=0,y index=1,col sep=space] {plots/primal/N20/drag_HLBM.dat};
        \addplot [line width=1pt,solid,color=green] table[x index=0,y index=1,col sep=space] {plots/primal/N20/dragBounceBack.dat};
        \addplot [line width=1pt,solid,color=orange] table[x index=0,y index=1,col sep=space] {plots/primal/N20/drag.dat};
        \addplot[solid,line width=2pt] coordinates {(15,5.58) (16,5.58)};
        \legend{HLBM, BGK+BounceBack, BGK+Bouzidi, Schäfer \textit{et al.}~\cite{schafer1996benchmark}}
      \end{axis}
    \end{tikzpicture}
    \caption[]{Comparison of the drag force over time for different incorporation of the solid cylinder in the flow. The cylinder diameter is resolved by $N=20$.}
    \label{fig:drag}
    \end{subfigure}
    \hfill
    \begin{subfigure}[t]{.49\textwidth}
    \begin{tikzpicture}
      \begin{axis}[
          width=\linewidth*1., % Scale the plot to \linewidth
          height=\linewidth*0.6,
          xmode=log,
          ymode=log,
          xlabel=$N$, % Set the labels
          ylabel=$\left|\frac{c_D-c_D^*}{c_D^*}\right|$,
          xlabel style={scale=1},
          ylabel style={scale=1},
          %x unit=\si{\degree}, % S   return util::pow(output, 1. / P);et the respective units
          %y unit=\si{\pascal},
          legend cell align=left,
          legend style={column sep=0.3cm,nodes={transform shape},at={(1,0)},anchor=south east}, % Put the legend below the plot
          x tick label style={anchor=north}, % Display labels sideways
          y tick label style={},
          xtick={20,40,80},
          xticklabels={20,40,80},
          grid
        ]
        \addplot[dashed,line width=1pt] coordinates {(20,0.1) (80,0.025)};
        \addplot [line width=1pt,solid,color=blue,mark=o] coordinates {(20, 0.03073820846)(40, 7.320e-3) (80, 2.064e-3)};
        \addplot [line width=1pt,solid,color=green,mark=o] coordinates {(20, 0.0387258) (40, 0.01271) (80, 5.719e-3)};
        \addplot [line width=1pt,solid,color=orange,mark=o] coordinates {(20, 9.337146057e-3)(40, 2.027e-3)(80, 5.104e-4)};
        \addplot[dashed,line width=1pt] coordinates {(20,0.1) (80,0.025)};
        \node[] at (60,7e-2) {\footnotesize$slope = -1$};
        \addplot[dashed,line width=1pt] coordinates {(20,5e-3) (80,3.125e-4)};
        \node[] at (25,1.5e-3) {\footnotesize$slope = -2$};
        %\legend{HLBM, HLBMBGK+BounceBack, BGK+Bouzidi, reference}
      \end{axis}
    \end{tikzpicture}
    \caption[]{Grid convergence study of the drag force for the different approaches. The relative error of the final drag force at $t = 16$ compared to the reference force from Schäfer \textit{et al.}~\cite{schafer1996benchmark} has been computed for increasing resolutions of $N = \{20, 40, 80\}$.}
    \label{fig:drag_eoc}
    \end{subfigure}
    \caption{Validation of the drag force for the cylinder modeled as a porous medium by HLBM.}
    \label{fig:primal_validation}
  \end{center}
\end{figure}

\noindent Fig.~\ref{fig:primal_validation} presents the results of the validation study of the primal model.
In Fig.~\ref{fig:drag}, the drag force is shown over the simulation duration for the different models.
All three solutions approach the reference value of $c_D^* = 5.58$ by reaching a steady level at $t = 16$s.
To ensure that the results are mesh-independent, a grid convergence study is conducted, in which the experimental order of convergence (EOC) has been evaluated in Fig.~\ref{fig:drag_eoc}.
Therein, the HLBM model achieves second-order convergence regarding the drag coefficient as the interpolated bounce back case.
The accuracy of the HLBM model lies between the bounce back and the interpolated bounce back method.

\subsection{Validation of the adjoint sensitivities: Steady laminar free flow around cylinder}\label{sec:adjointValidation}

Next, the free flow around a cylinder is simulated for the 2D and 3D cases in the steady laminar regime.
The adjoint sensitivities are computed and validated against gradients computed via finite difference quotients (FDQ) and forward AD.

\subsubsection{2D case at $Re = 20$}\label{sec:adjointValidation2D}

\begin{figure}[h!]
    \centering
    \includegraphics[width=0.6\linewidth]{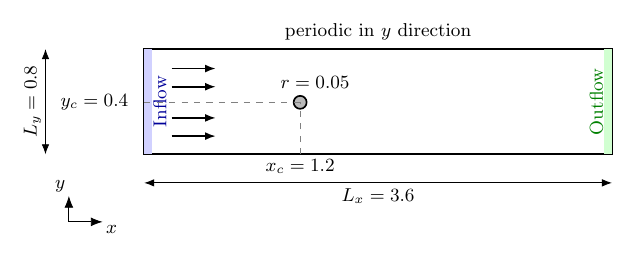}
    \caption{Schematic illustration of the simulation setup in 2D with the used domain sizes and cylinder positions.}
    \label{fig:setup2d}
\end{figure}

The 2D free flow around a cylinder, as illustrated in Fig.~\ref{fig:setup2d}, is investigated.
The flow domain spans over $L_x \times L_y$ and the cylinder center is positioned at $(x_c, y_c)$ with a cylinder radius of $r$.
At the inflow, a Dirichlet velocity boundary condition with $\bm{u}_\infty = (0.2, 0.0)^\transpose$ m/s and at the outlet a Dirichlet pressure boundary condition with $p_{\mathrm{out}} = 0$ is applied.
On the lateral walls, periodic boundary conditions are applied.

\noindent The Reynolds number is set to $Re = 20$.
All relevant simulation parameters on the simulation domain and parameters regarding the setup are summarized in Tab.~\ref{tab:adjointValidation2D}.

\begin{table}[h!]
\centering
\caption{Simulation parameters for the produced results in Sec.~\ref{sec:adjointValidation2D}.\label{tab:adjointValidation2D}}
\begin{threeparttable}
\begin{tabular}{c|cc}
\hline \hline 
\textbf{Parameter} & \textbf{Value} & \textbf{Unit}\\
\hline
Domain extent ($L_x$, $L_y$) & (3.6, 0.8) & m\\
Cylinder center ($x_c$, $y_c$) & (1.2, 0.4) & m\\
Radius cylinder ($r$) & 0.05 & m\\
Resolution over cylinder diameter ($N$) & 21 & -\\
Lattice relaxation time ($\tau$) & 0.55 & -\\
Inflow velocity ($\bm{u}_\infty$) & (0.2, 0.0) & m/s\\
Permeability of cylinder ($K$) & $10^{-6}$ & $\mathrm{m}^2$ \\
Fluid density ($\rho$) & 1 & kg/$\mathrm{m}^3$\\
Max simulation time ($T$) & 32 & s\\
Center of $\Gamma_{\triangle x, J}$ for $J_{\mathrm{FFA}}$ & (1.7, 0.4) & m\\
Extent of $\Gamma_{\triangle x, J}$ for $J_{\mathrm{FFA}}$ & (0.00476, 0.8) & m\\
Center of $\Omega_{\triangle x, J}$ for $J_{\mathrm{DISS}}$ & (1.2, 0.4) & m\\
Extent of $\Omega_{\triangle x, J}$ for $J_{\mathrm{DISS}}$ & (2.2, 0.8) & m\\
\hline \hline
\end{tabular}
\end{threeparttable}
\end{table}

\begin{figure}[h]
    \centering
    \begin{subfigure}[b]{.65\textwidth}
    \raisebox{0.mm}{\includegraphics[width=1.\linewidth]{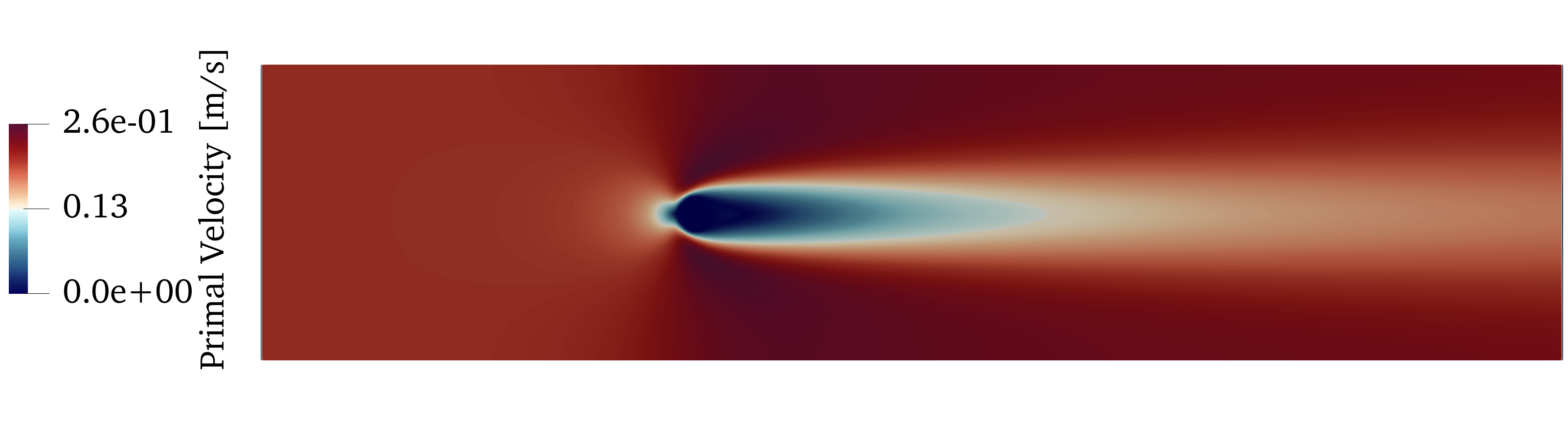}}
    \caption{Primal velocity field.}
    \label{fig:primalVelocity}
    \end{subfigure}
    \hfill
    \begin{subfigure}[b]{.33\textwidth}
    \raisebox{0.mm}{\includegraphics[width=0.55\linewidth]{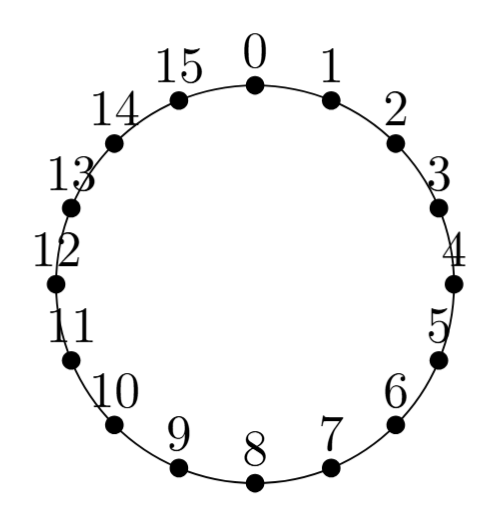}}
    \caption{Sample positions for the gradient comparison.}
    \label{fig:samplePositions}
    \end{subfigure}
    
    \begin{subfigure}[b]{.45\textwidth}
    \raisebox{0.mm}{\includegraphics[width=1.\linewidth]{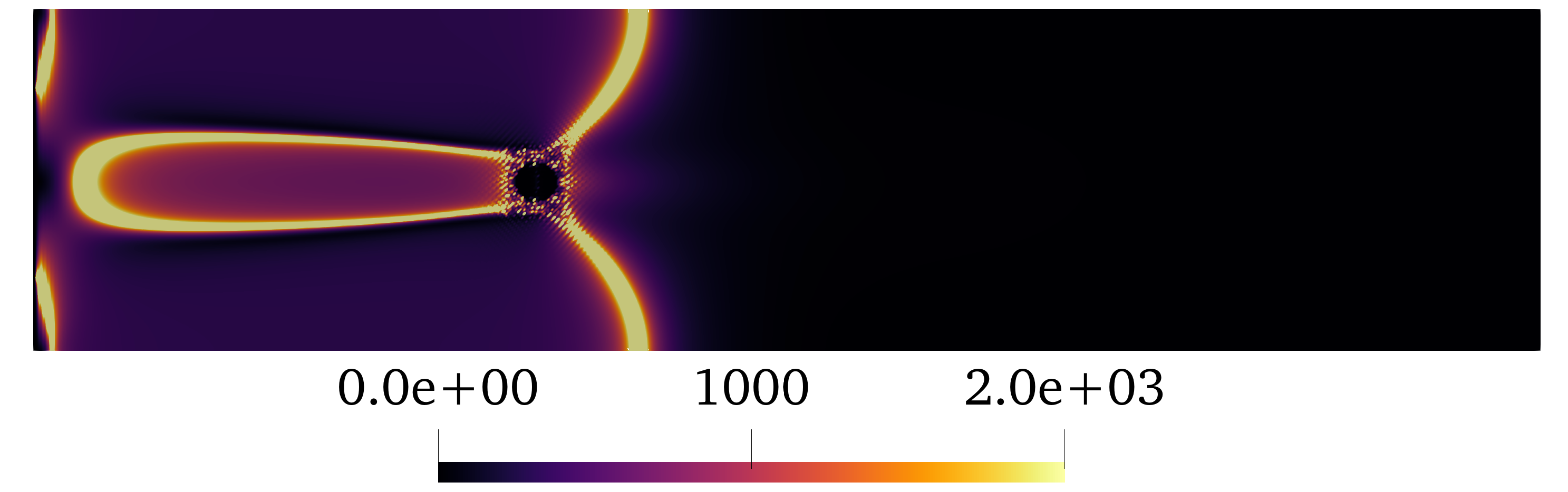}}
    \caption{Adjoint velocity field for the drag MEA objective.}
    \label{fig:adjointVelocityMEA}
    \end{subfigure}
    \hfill
    \begin{subfigure}[b]{.54\textwidth}
    \raisebox{2.mm}{\includegraphics[width=1.\linewidth]{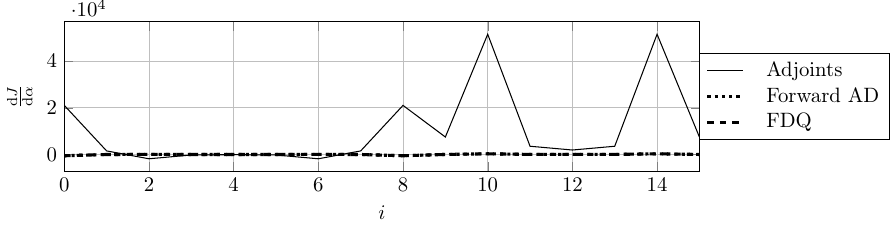}}
    \caption{Surface gradient comparison.}
    \label{plot:gradMEA}
    \end{subfigure}
    
    \begin{subfigure}[b]{.45\textwidth}
    \raisebox{0.mm}{\includegraphics[width=1.\linewidth]{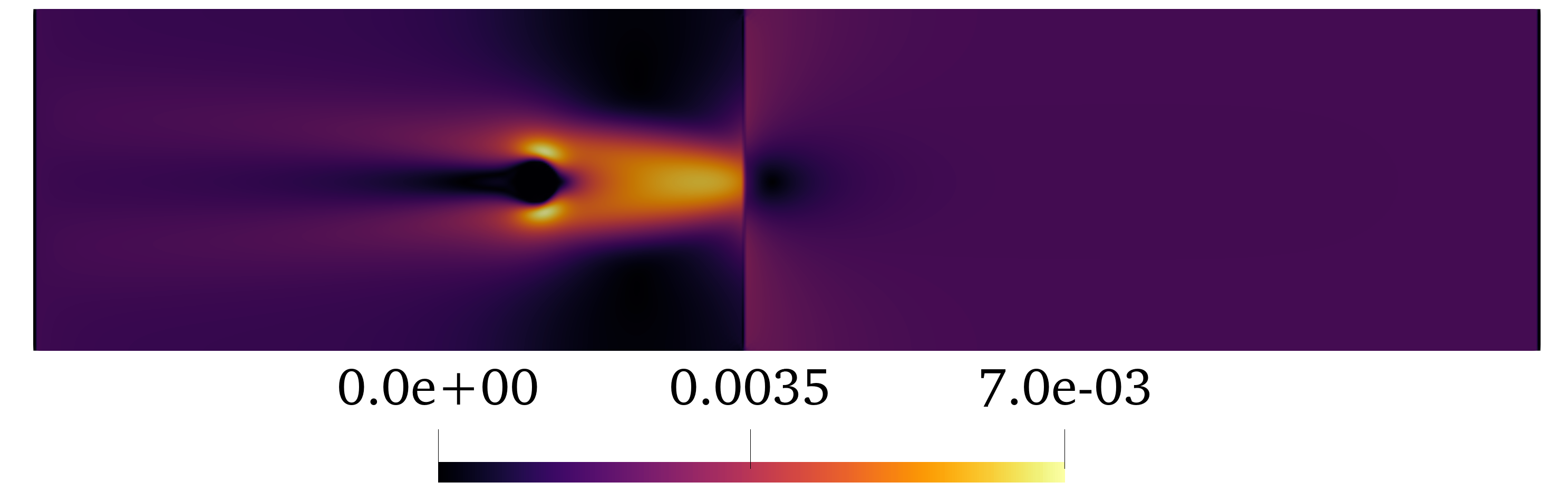}}
    \caption{Adjoint velocity field for the drag FFA objective.}
    \label{fig:adjointVelocityFFA}
    \end{subfigure}
    \hfill
    \begin{subfigure}[b]{.54\textwidth}
    \raisebox{2.mm}{\includegraphics[width=1.\linewidth]{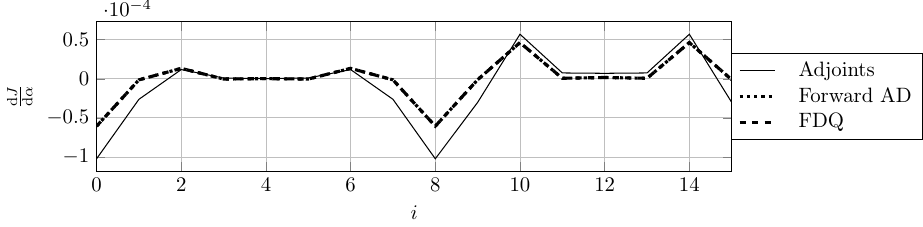}}
    \caption{Surface gradient comparison.}
    \label{plot:gradFFA}
    \end{subfigure}
    
    \begin{subfigure}[b]{.45\textwidth}
    \raisebox{0.mm}{\includegraphics[width=1.\linewidth]{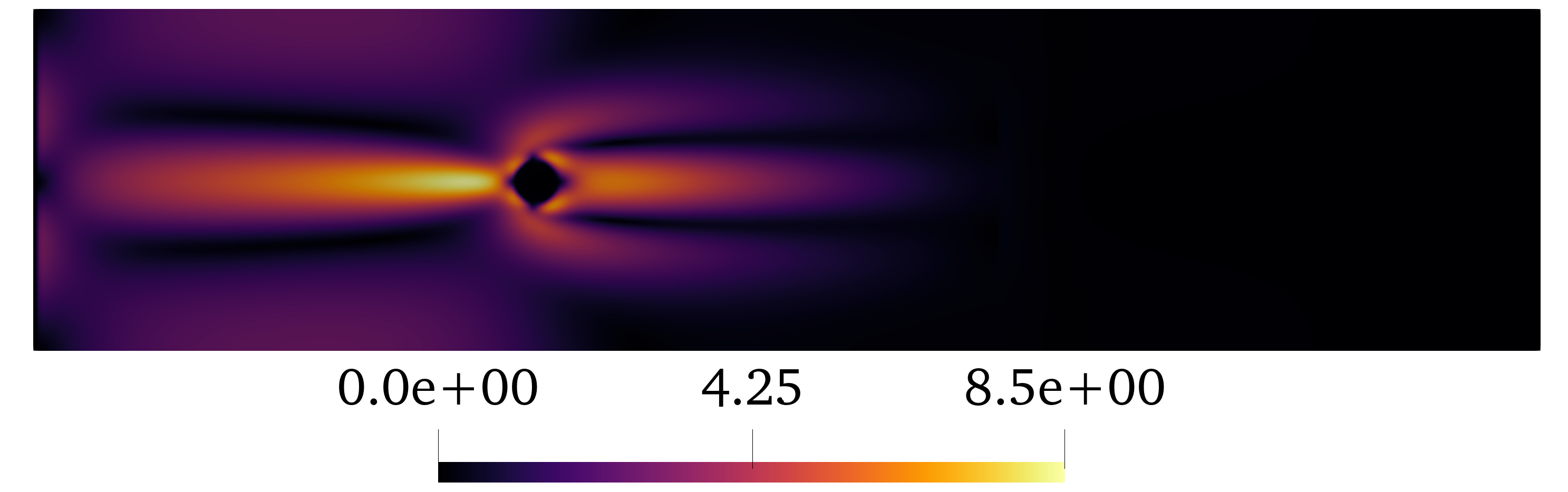}}
    \caption{Adjoint velocity field for the dissipation objective.}
    \label{fig:adjointVelocityDISS}
    \end{subfigure}
    \hfill
    \begin{subfigure}[b]{.54\textwidth}
    \raisebox{2.mm}{\includegraphics[width=1.\linewidth]{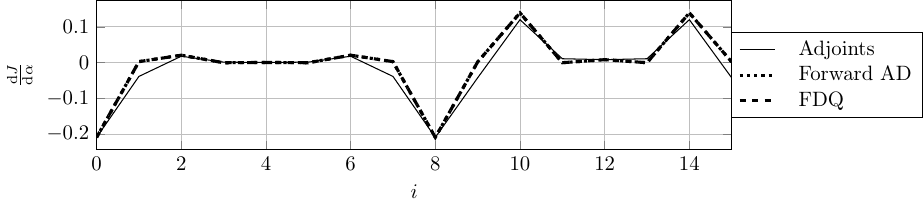}}
    \caption{Surface gradient comparison.}
    \label{plot:gradDISS}
    \end{subfigure}
    \caption{Validation of the adjoint gradients for the different objective functionals. The surface gradients are compared for adjoint-based, forward AD, and FDQ.}
    \label{fig:2d_adjointValidation}
\end{figure}

The primal velocity distribution is shown in Fig.~\ref{fig:primalVelocity}.
For the validation of the adjoint gradients, the sample points illustrated in Fig.~\ref{fig:samplePositions} on the cylinder surface are chosen, on which additionally FDQ and forward AD sensitivities are computed.
Fig.~\ref{fig:adjointVelocityMEA}, Fig.~\ref{fig:adjointVelocityFFA}, and Fig.~\ref{fig:adjointVelocityDISS} picture the adjoint velocity distribution with the corresponding gradient comparison in Fig.~\ref{plot:gradMEA}, Fig.~\ref{plot:gradFFA}, and Fig.~\ref{plot:gradDISS} for the three objective functionals~\eqref{eq:obj_mea}, \eqref{eq:obj_ffa}, and \eqref{eq:obj_diss}, respectively.
In the adjoint simulation for the drag objective functional based on the MEA, we observe instabilities in the cylinder wall vicinity as well as deviating adjoint gradients from those computed by the other two approaches.
This coincides with the observations made in the work of Stück \textit{et al.}~\cite{stuck2012adjoint} that the control domain and objective domain for controlled body shapes in a free flow setting should be chosen in a way that they are not only defined on the body surface.
Instead, the control domain and the objective domain should be separated, leading to fewer instabilities and smoother adjoint solutions as we observe in Fig.~\ref{fig:adjointVelocityFFA} and Fig.~\ref{fig:adjointVelocityDISS}.
For the drag objective based on the FFA and the dissipation objective, the adjoint gradients match the sensitivities obtained from FDQ and forward AD, except for small deviations around the sample positions close to the flow separation points.
Those deviations align with the results in the work of Cheylan \textit{et al.}~\cite{cheylan_shape_2019} who refer to prior works of Nadarajah~\cite{nadarajah2003discrete} stating that this observation is commonly reported in adjoint cylinder flow literature.
In the following numerical experiments, we focus on the drag objective based on the FFA and the dissipation objective.

\subsubsection{3D case at $Re = 40$}\label{sec:adjointValidation3D}

\begin{figure}[h]
    \centering
    \includegraphics[width=0.6\linewidth]{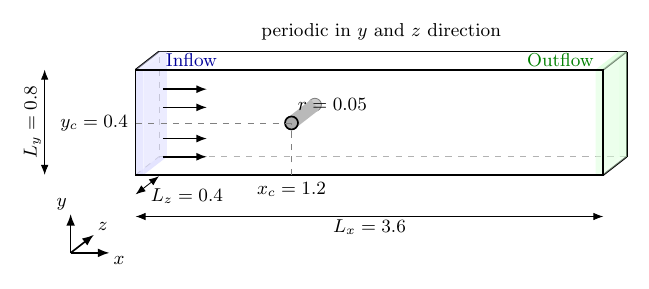}
    \caption{Schematic illustration of the simulation setup in 3D with the used domain sizes and cylinder positions.}
    \label{fig:setup3d}
\end{figure}

Now the 3D case of~\ref{sec:adjointValidation2D} is investigated for a slightly higher Reynolds number of $Re = 40.$
To save computation time for the 3D cases, the simulations are conducted with single floating-point precision.
The influence of the floating point precision has been evaluated and showed no visible difference for the validation results, cf.~\ref{app:FPP}.
Fig.~\ref{fig:setup3d} illustrates the simulation setup for the 3D case.

\noindent Along the z-axis, periodic boundary conditions are applied.
For all the other boundaries, the identical boundary treatment is applied as in the 2D case.
Again, all relevant simulation parameters and parameters regarding the setup are listed in Tab.~\ref{tab:adjointValidation3D}.

\begin{table}[h!]
\centering
\caption{Simulation parameters for the produced results in Sec.~\ref{sec:adjointValidation3D}.\label{tab:adjointValidation3D}}
\begin{threeparttable}
\begin{tabular}{c|cc}
\hline \hline 
\textbf{Parameter} & \textbf{Value} & \textbf{Unit}\\
\hline
Domain extent ($L_x$, $L_y$, $L_z$) & (3.6, 0.8, 0.4) & m\\
Cylinder center axis along ($x_c$, $y_c$) & (1.2, 0.4) & m\\
Radius cylinder ($r$) & 0.05 & m\\
Resolution over cylinder diameter ($N$) & 21 & -\\
Lattice relaxation time ($\tau$) & 0.55 & -\\
Inflow velocity ($\bm{u}_\infty$) & (0.2, 0.0, 0.0) & m/s\\
Permeability of cylinder ($K$) & $10^{-6}$ & $\mathrm{m}^2$ \\
Fluid density ($\rho$) & 1 & kg/$\mathrm{m}^3$\\
Max simulation time ($T$) & 32 & s\\
Center of $\Gamma_{\triangle x, J}$ for $J_{\mathrm{FFA}}$ & (1.7, 0.4, 0.2) & m\\
Extent of $\Gamma_{\triangle x, J}$ for $J_{\mathrm{FFA}}$ & (0.00476, 0.8, 0.4) & m\\
Center of $\Omega_{\triangle x, J}$ for $J_{\mathrm{DISS}}$ & (1.2, 0.4, 0.2) & m\\
Extent of $\Omega_{\triangle x, J}$ for $J_{\mathrm{DISS}}$ & (2.2, 0.8, 0.4) & m\\
\hline \hline
\end{tabular}
\end{threeparttable}
\end{table}

\begin{figure}
    \centering
    \begin{subfigure}[b]{.49\textwidth}
    \raisebox{0.mm}{\includegraphics[width=1.\linewidth]{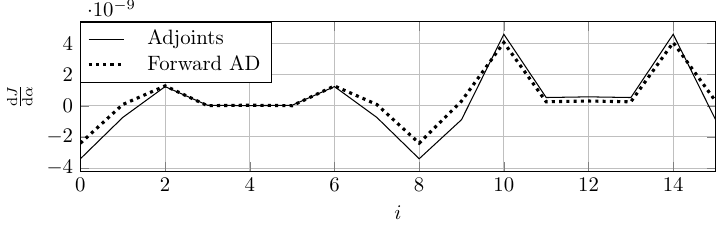}}
    \caption{Surface gradient comparison for the FFA objective.}
    \label{plot:gradFFA3d}
    \end{subfigure}
    \hfill
    \begin{subfigure}[b]{.49\textwidth}
    \raisebox{0.mm}{\includegraphics[width=1.\linewidth]{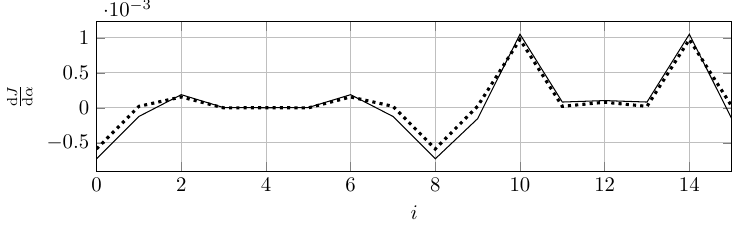}}
    \caption{Surface gradient comparison for the dissipation objective.}
    \label{plot:gradDISS3d}
    \end{subfigure}
    \caption{Validation of the adjoint gradients for the different objective functionals. The surface gradients are compared for adjoint-based, forward AD, and FDQ. In the 3D case, the simulation has been conducted with single floating-point precision to save computational resources.}
    \label{fig:adjointValidation3d}
\end{figure}

The validation of the adjoint sensitivities is demonstrated in Fig.~\ref{fig:adjointValidation3d}.
The same sample point distribution as shown in Fig.~\ref{fig:samplePositions} is used on the x-y-plane at $z = 0.2$.
For the FFA and dissipation objective, the adjoint gradients have a similar accuracy as in the 2D case.
Similarly, there are slight discrepancies between the adjoint and forward AD gradients for the points close to the top and bottom sides of the cylinder, but for the remaining positions, the gradients match very well.

\subsection{Mean-flow adjoint sensitivities: Unsteady laminar free flow around cylinder}\label{sec:meanFlowAdjoints}

In this section, the laminar unsteady case for the 2D cylinder flow at $Re = 100$ is investigated.
This well-studied case is known to develop the von Kármán vortex street triggered by the Helmholtz instabilities occurring in the wake of the cylinder~\cite{abernathy1962formation}.
The simulation setup is identical to that illustrated in Fig.~\ref{fig:setup2d}.
Besides the Reynolds number and the maximal simulation time of $T = 300$s, all other simulation parameters remain identical as given in Tab.~\ref{tab:adjointValidation2D}.
The tracking average of the particle distribution function, objective functional, and the derivatives of the objective functionals required to evaluate \eqref{eq:averageObjective}, \eqref{eq:adjoint_unsteady}, and \eqref{eq:gradients_unsteady}, are computed starting after $80\%$ of the maximum simulation time is passed.

\begin{figure}[h]
    \centering
    \begin{subfigure}[b]{.49\textwidth}
    \raisebox{0.mm}{\includegraphics[width=1.\linewidth]{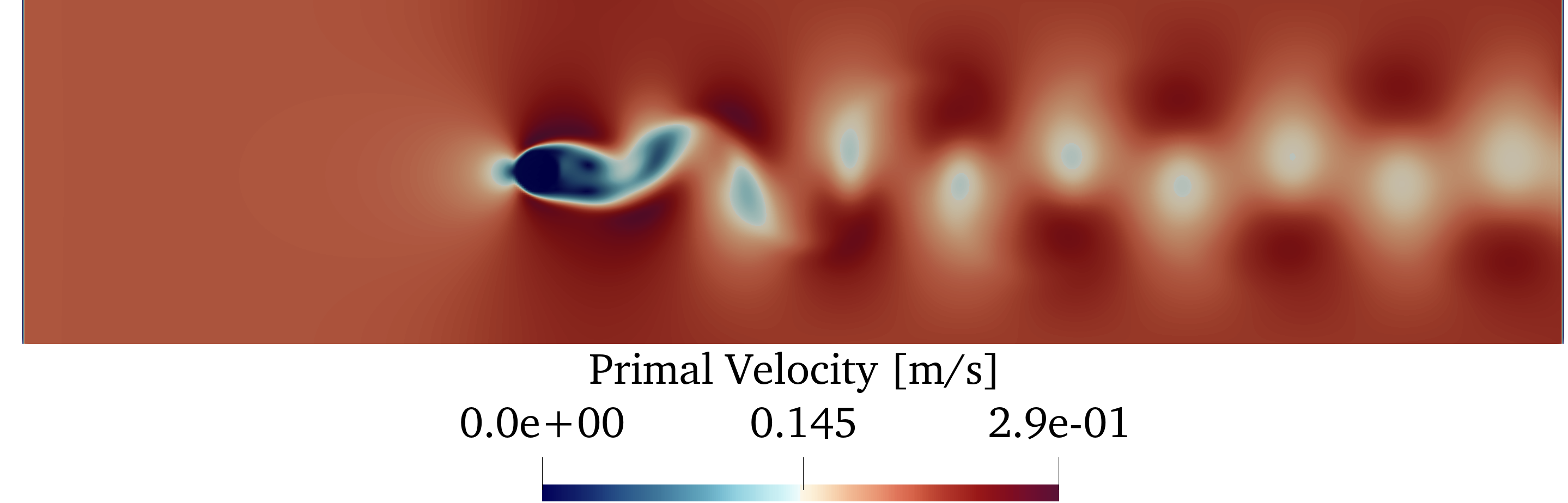}}
    \caption{Snapshot of the primal velocity field.}
    \label{fig:primalVelocityUnsteady}
    \end{subfigure}
    \hfill
    \begin{subfigure}[b]{.49\textwidth}
    \raisebox{0.mm}{\includegraphics[width=1.\linewidth]{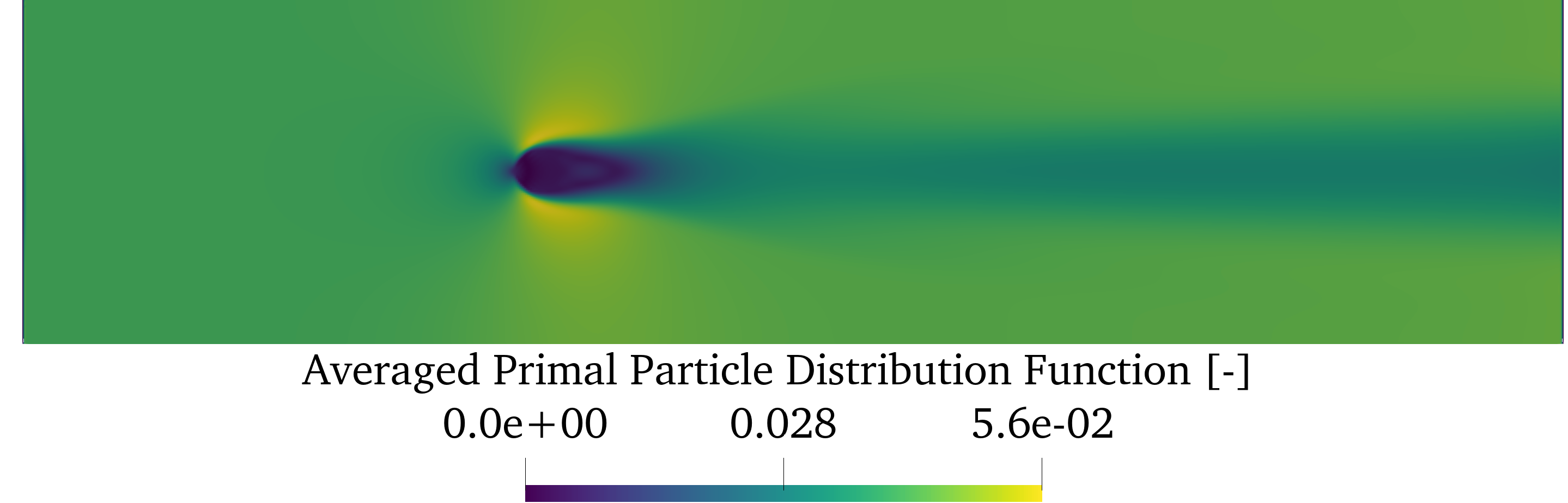}}
    \caption{Averaged distribution of the particle distribution function.}
    \label{fig:primalPopulationsUnsteady}
    \end{subfigure}
    
    \begin{subfigure}[b]{.49\textwidth}
    \raisebox{0.mm}{\includegraphics[width=1.\linewidth]{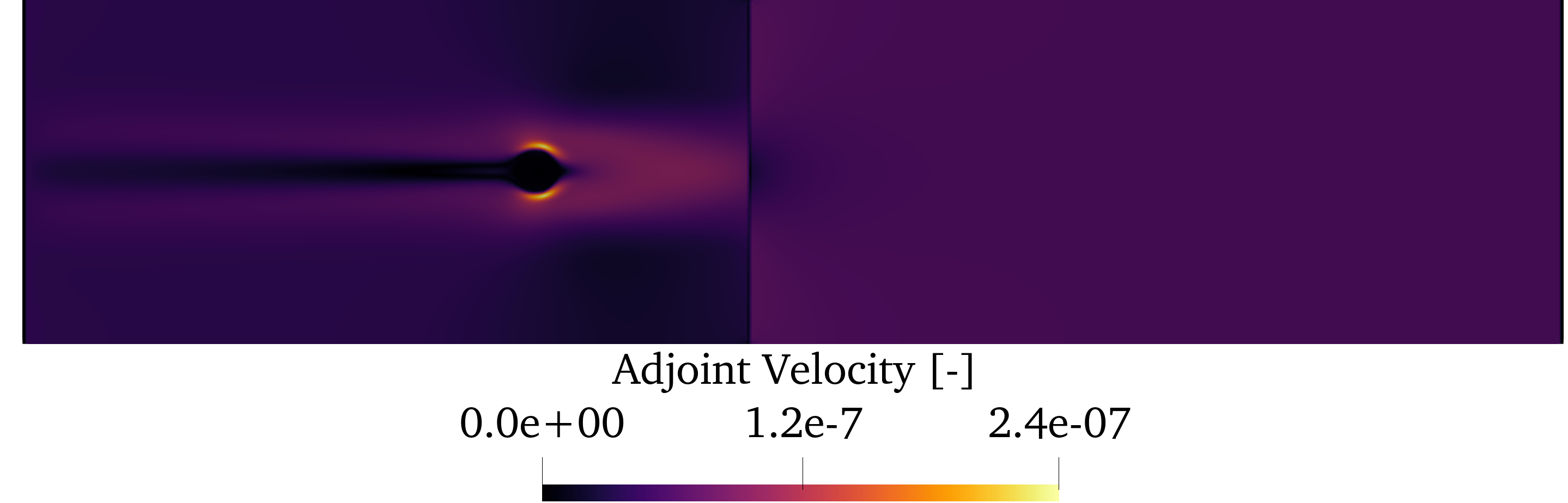}}
    \caption{Adjoint velocity field for the drag FFA objective.}
    \label{fig:adjointVelocityUnsteadyFFA}
    \end{subfigure}
    \hfill
    \begin{subfigure}[b]{.49\textwidth}
    \raisebox{0.mm}{\includegraphics[width=1.\linewidth]{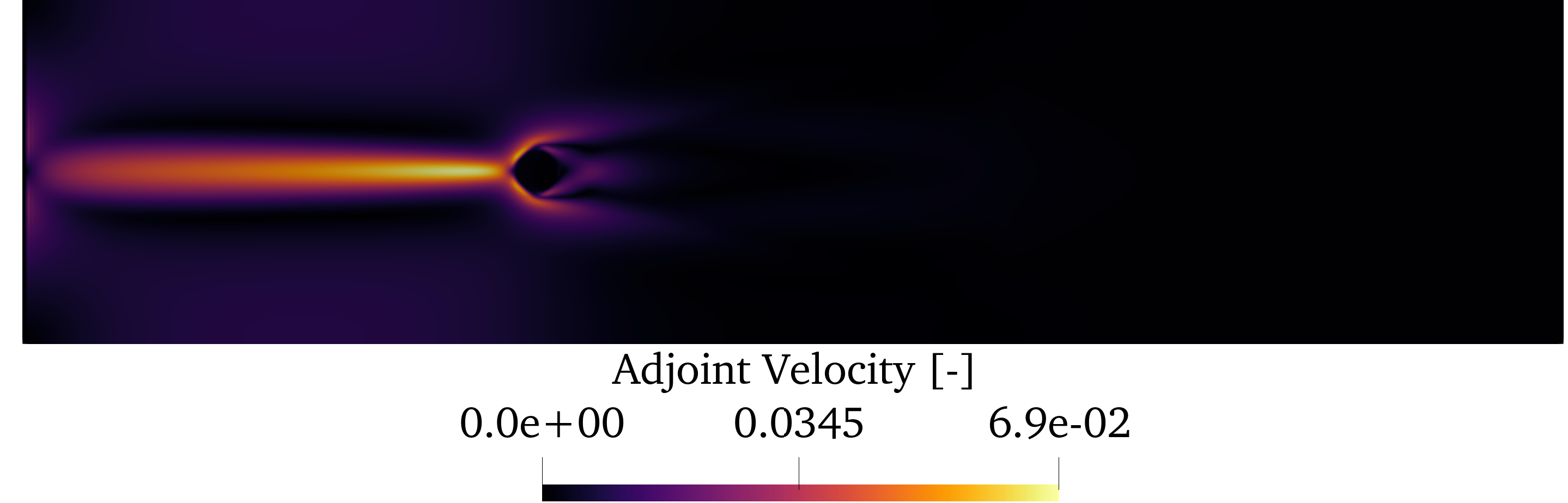}}
    \caption{Adjoint velocity field for the dissipation objective.}
    \label{fig:adjointVelocityUnsteadyDISS}
    \end{subfigure}
    
    \begin{subfigure}[b]{.49\textwidth}
    \raisebox{0.mm}{\includegraphics[width=1.\linewidth]{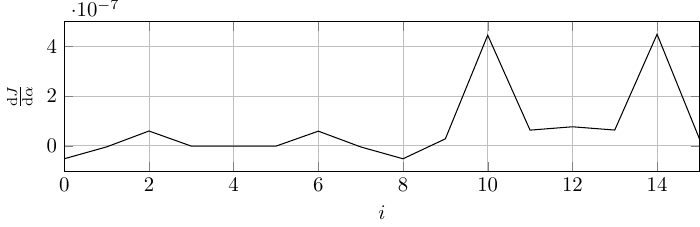}}
    \caption{Surface gradient comparison for the FFA objective.}
    \label{plot:gradFFAUnsteady}
    \end{subfigure}
    \hfill
    \begin{subfigure}[b]{.49\textwidth}
    \raisebox{0.mm}{\includegraphics[width=1.\linewidth]{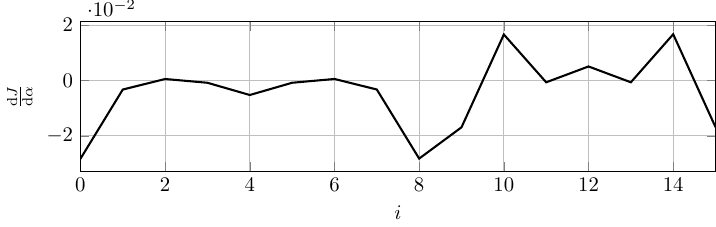}}
    \caption{Surface gradient comparison for the dissipation objective.}
    \label{plot:gradDISSUnsteady}
    \end{subfigure}
    \caption{Mean-flow adjoint of unsteady laminar 2D cylinder flow at $Re = 100$.}
    \label{fig:2d_meanFlowAdjoint}
\end{figure}

Fig.~\ref{fig:2d_meanFlowAdjoint} shows the results for the laminar unsteady case.
In Fig.~\ref{fig:primalVelocityUnsteady}, a snapshot of the primal velocity field is visualized, where the fully-developed von Kármán vortex street can be observed.
This oscillating nature enforces this case to be unsteady while being in the fully laminar regime.
Fig.~\ref{fig:primalPopulationsUnsteady} shows the averaged distribution of the particle distribution function at the end of the primal simulation.
Therein, the vortex street is smoothed out, and the symmetrical distribution along the axis $y = 0.2$m of the flow domain indicates a sufficiently long averaging time interval to remove any influences or bias due to the unsteady fluctuations.
Fig.~\ref{fig:adjointVelocityUnsteadyFFA} shows the mean-flow adjoint velocity distribution. 
Compared to the steady laminar case in Fig.~\ref{fig:adjointVelocityFFA}, high adjoint velocities are concentrated in the vicinity of the cylinder with a more pronounced "wake" in the upstream direction.
For the dissipation objective in Fig.~\ref{fig:adjointVelocityUnsteadyDISS}, similar observations can be made.
The jet upstream of the cylinder is now much more prominent compared to the laminar case.
Furthermore, the main activity of the adjoint velocity happens upstream of the cylinder compared to Fig.~\ref{fig:adjointVelocityDISS}, indicating the sensitive region in the flow regarding the objective.
The adjoint velocity distribution shows now higher spatial gradients for this case with a higher Reynolds number, and as the adjoint problem is fed only by the mean-flow primal solution, the adjoint state field is again symmetrical.
The surface gradients are sampled again on the positions given in Fig.~\ref{fig:samplePositions} and show symmetrical distributions for both objective functionals.
The obtained surface gradients and adjoint velocity distributions based on the mean-flow adjoint match the results observed in the work of Cheylan \textit{et al.}~\cite{cheylan_shape_2019} very well.
Note that for this unsteady laminar flow, the mean-flow adjoint-based sensitivities yielded stable results, while those computed via forward AD had serious limitations in the gradient accuracy as reported in~\cite{LSS2018, Wang_Gao_2013}.
The observations on this issue are presented in~\ref{appendix:forward_ad_divergence}, where the gradients based on forward AD were not feasible for practical use due to bloating of the sensitivities caused by the Helmholtz instability and accumulating gradient errors due to diverging solution trajectories. 
Those results, together with those presented in Sec.~\ref{sec:adjointValidation2D}, Sec.~\ref{sec:adjointValidation3D}, and the current section, demonstrate that the mean-flow-based adjoint approach is a practical solution compromising numerical efficiency, memory footprint, stability of the computed gradients, and their accuracy. 

\subsection{Proof-of-concept: Turbulent free flow around cylinder}\label{sec:proof-of-concept}

To further explore the capabilities and extendibility of the current approach, the 3D turbulent cylinder flow at $Re = 3900$ is computed with the mean-flow adjoint approach.
The 3D cylinder flow at $Re = 3900$ is an established LES benchmark case, where the turbulent flow around the cylinder is in the transient regime~\cite{LES2016, FRANKE20021191}.
To the author's knowledge, it remains an open question if accurate gradients can be computed for turbulent flows with chaotic solutions with respect to the initial conditions.
It is generally acknowledged that the gradients become inaccurate due to the fact that, by the chaotic features of the flow, the trajectories of the solution state diverge over time~\cite{LSS2018, Wang_Gao_2013}.
Mathematically, this can be quantified by the Lyapunov exponent, e.g., if the exponent exceeds one for any dynamic system, the sensitivities computed by conventional methods diverge~\cite{Wang_Gao_2013}.
It is not the current scope of the present work to assess what is possible to retrieve accurate gradients for turbulent flows in general, but we will compare the obtained surface gradients for a fully differentiated LES model (here HRRLBM-LES) and the gradients computed via the FTA, where the turbulence model is ignored in the adjoint simulation.
The FTA corresponds to setting $\partial \tau_{\mathrm{turb}} / \partial f_i = 0$~\cite{MARTA2013102} and uses the averaged value for $\tau_{\mathrm{turb}}$ as a constant in the adjoint HRRLBM collision operator without the differentiated LES model.
Moreover, it is also not the scope of the current paper to deliver dedicated turbulence statistics evaluation for the current cylinder flow, as the primal model has been validated in the turbulent regime against experimental results in several previous works~\cite{kummerlander_efficient_vocal_2026, kummerlander_efficient_2026-1, coreixas2017recursive, jacob2018new, teutscher_digital_2025}.

The simulation setup for the turbulent case is identical to Sec.~\ref{sec:adjointValidation3D} as sketched in Fig.~\ref{fig:setup3d} with slightly different simulation parameters as listed in Tab.~\ref{tab:proof-of-concept}.
The averaging window is set between $[100, 800]$s to track the averaged entities required for the evaluation of the adjoint Jacobian expressions.
The drag functional based on the FFA is used for the adjoint simulations.

\begin{table}[h!]
\centering
\caption{Simulation parameters for the produced results in Sec.~\ref{sec:proof-of-concept}.\label{tab:proof-of-concept}}
\begin{threeparttable}
\begin{tabular}{c|cc}
\hline \hline 
\textbf{Parameter} & \textbf{Value} & \textbf{Unit}\\
\hline
Domain extent ($L_x$, $L_y$, $L_z$) & (4.0, 1.0, 0.4) & m\\
Cylinder center axis along ($x_c$, $y_c$) & (0.7, 0.5) & m\\
Radius cylinder ($r$) & 0.05 & m\\
Resolution over cylinder diameter ($N$) & 21 & -\\
Lattice relaxation time ($\tau$) & 0.501 & -\\
Inflow velocity ($\bm{u}_\infty$) & (0.2, 0.0, 0.0) & m/s\\
Permeability of cylinder ($K$) & $3.85\times10^{-8}$ & $\mathrm{m}^2$ \\
Fluid density ($\rho$) & 1 & kg/$\mathrm{m} ^3$\\
Max simulation time ($T$) & 800 & s\\
Smagorinsky constant ($C_s$) & 0.1 & s\\
Center of $\Gamma_{\triangle x, J}$ for $J_{\mathrm{FFA}}$ & (1.2, 0.5, 0.2) & m\\
Extent of $\Gamma_{\triangle x, J}$ for $J_{\mathrm{FFA}}$ & (0.004, 0.6, 0.4) & m\\
\hline \hline
\end{tabular}
\end{threeparttable}
\end{table}

\begin{figure}
    \centering
    \includegraphics[width=\linewidth]{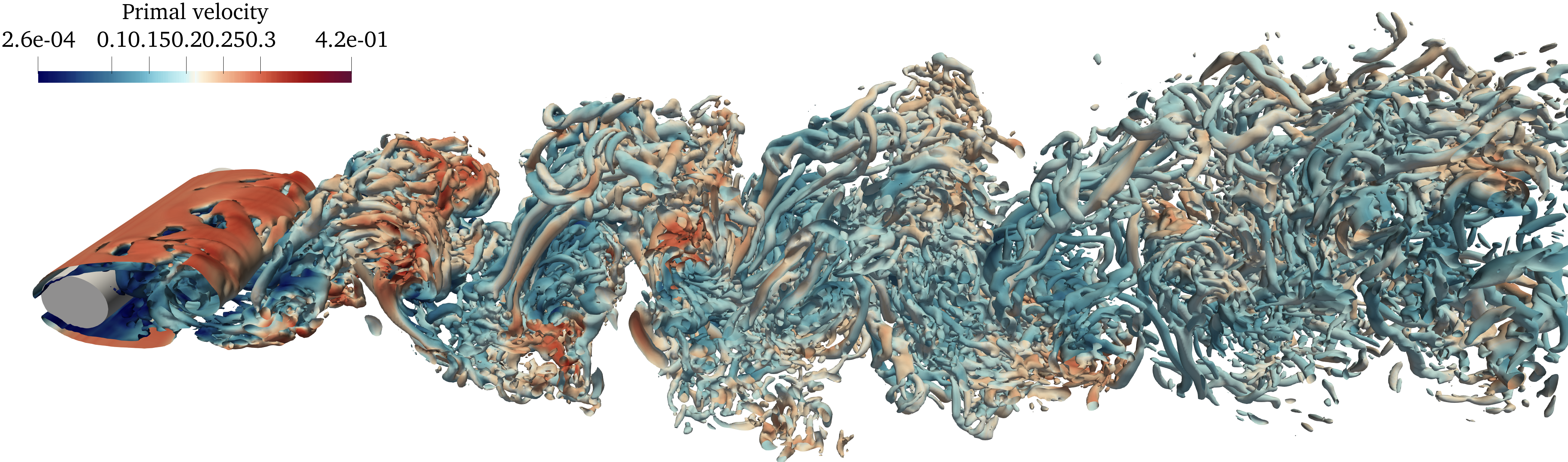}
    \caption{Contour plot of the Q-criterion at $Q = 2$ for the primal velocity.}
    \label{fig:qcriterion}
\end{figure}

\begin{figure}
    \centering
    \begin{subfigure}[b]{.49\textwidth}
    \raisebox{0.mm}{\includegraphics[width=1.\linewidth]{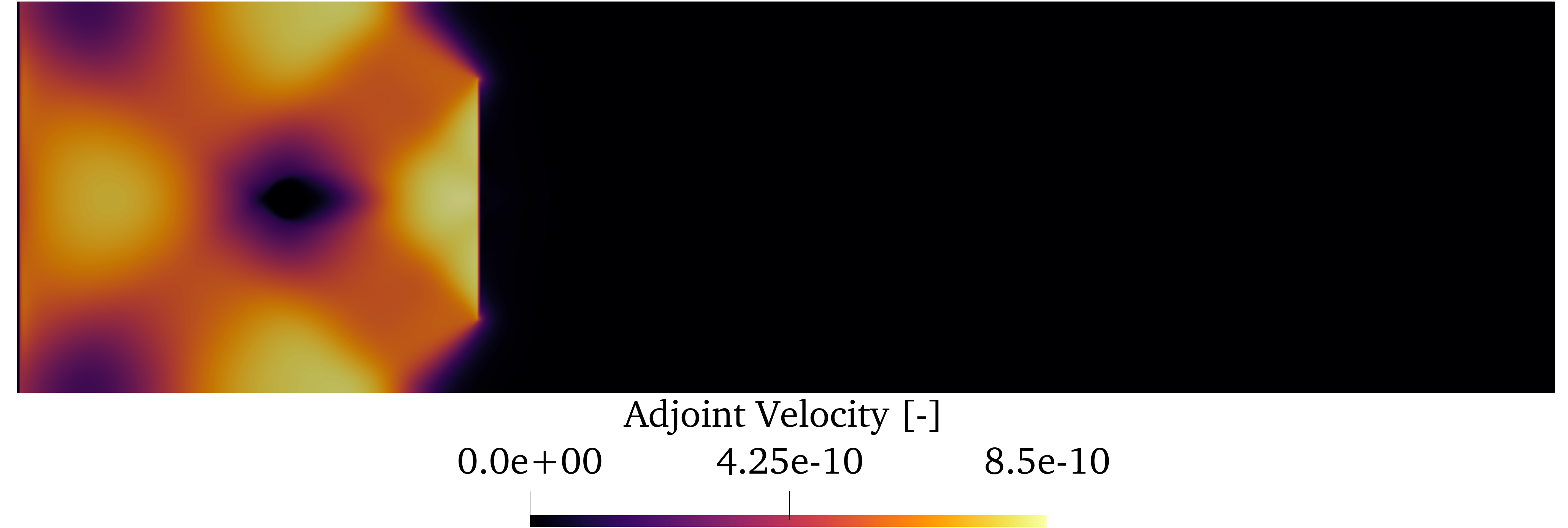}}
    \caption{Adjoint velocity field for the drag FFA objective.}
    \label{fig:adjointVelocityFFATurbulent}
    \end{subfigure}
    \hfill
    \begin{subfigure}[b]{.49\textwidth}
    \raisebox{0.mm}{\includegraphics[width=1.\linewidth]{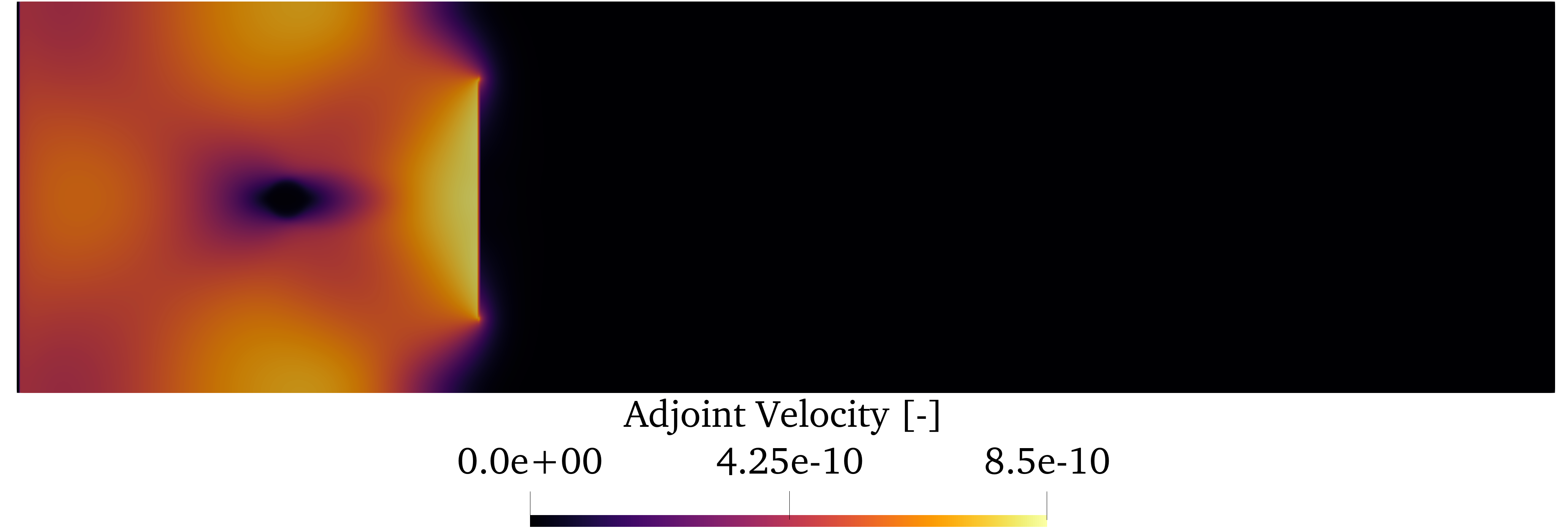}}
    \caption{Adjoint velocity field for the drag FFA objective with the FTA.}
    \label{fig:adjointVelocityFFATurbulentFTA}
    \end{subfigure}
    \caption{Comparison of the adjoint solutions for the adjoint HHRR-LES model and the FTA.}
    \label{fig:adjointVelocityTurbulent}
\end{figure}

Fig.~\ref{fig:qcriterion} visualizes the iso-surface for $Q = 2$ of the turbulent cylinder flow with the projected primal velocity magnitude.
As clearly visible, the flow is fully transient and chaotic, where conventional methods for gradient computation would fail according to~\cite{LSS2018}.
By employing the mean-flow-based adjoint approach, the adjoint velocity distributions are obtained and shown in Fig.~\ref{fig:adjointVelocityTurbulent}.
Therein, Fig.~\ref{fig:adjointVelocityFFATurbulent} shows the adjoint velocity field for the fully differentiated HRRLBM-LES model, and Fig.~\ref{fig:adjointVelocityFFATurbulentFTA} contains the solution with the FTA.
In both cases, the middle x-y plane has been visualized.
We can clearly observe that the adjoint velocity fields differ for the two cases and that in the solution with FTA, the distribution looks smoother compared to the fully differentiated case.
Especially, in the vicinity of the cylinder, the adjoint velocity magnitude is overall lower in the FTA case.
To achieve convergence in the adjoint solver for the highly turbulent regime, we employed a regularization approach that adds numerical diffusion in the adjoint simulation to suppress spurious oscillations originating in the high-shear region close to the cylinder.
This is a commonly applied measure to stabilize adjoint simulations in turbulent flows, as stated by~\cite{cheylan_shape_2019,blonigan2012towards}.
In the current work, this is realized by increasing the lattice relaxation time in the adjoint simulation, which has been set to $\tau_{\mathrm{adj}} = 0.6$ in both cases shown in Fig.~\ref{fig:adjointVelocityTurbulent}.

\begin{figure}[h!]
    \centering
    \begin{subfigure}[b]{.49\textwidth}
    \centering
    \raisebox{0.mm}{\includegraphics[width=0.56\linewidth]{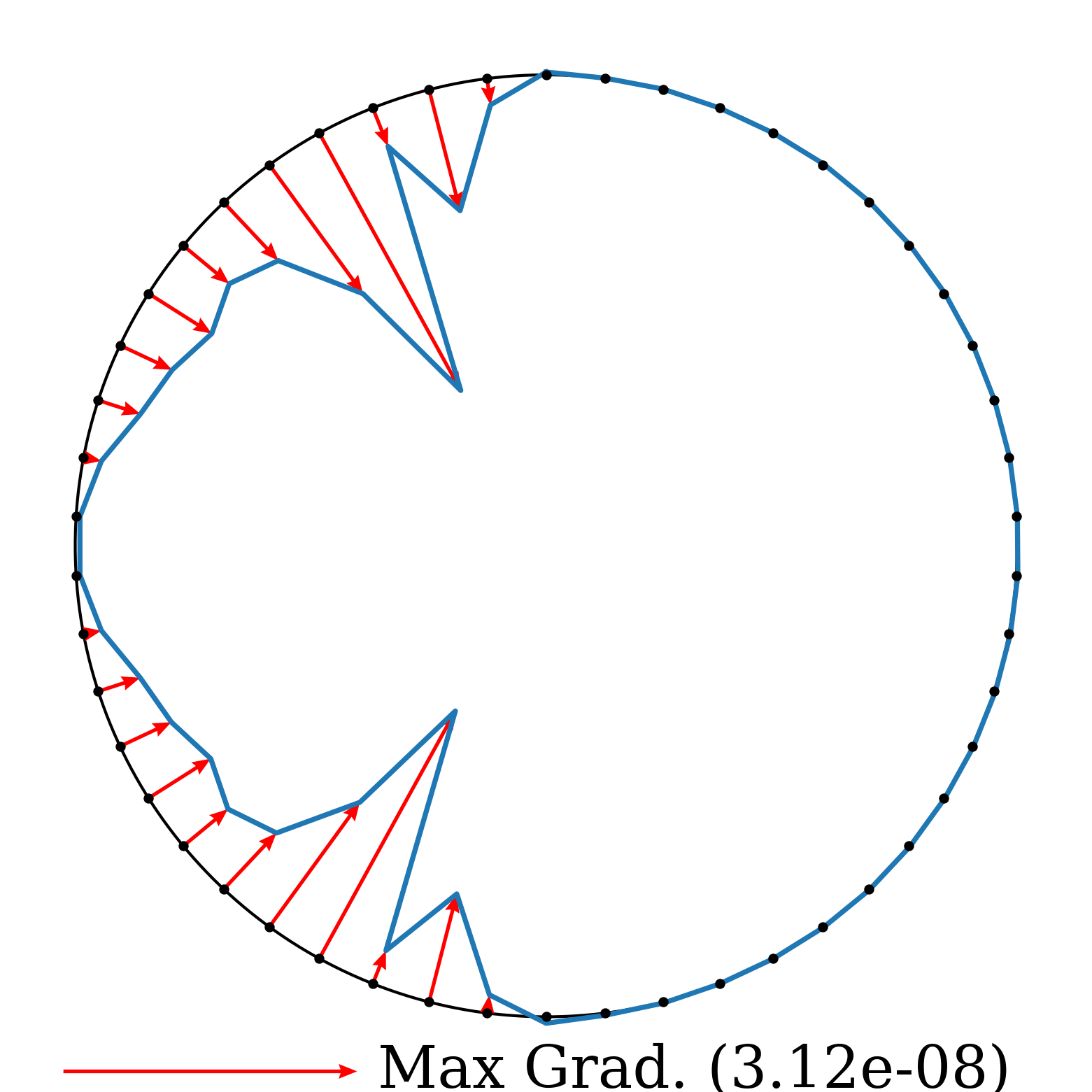}}
    \caption{Adjoint HRRLBM-LES model with $\tau_{\mathrm{adj}} = 0.6$.}
    \label{plot:surfaceARLBMtau0.6}
    \end{subfigure}
    \hfill
    \begin{subfigure}[b]{.49\textwidth}
    \centering
    \raisebox{0.mm}{\includegraphics[width=0.56\linewidth]{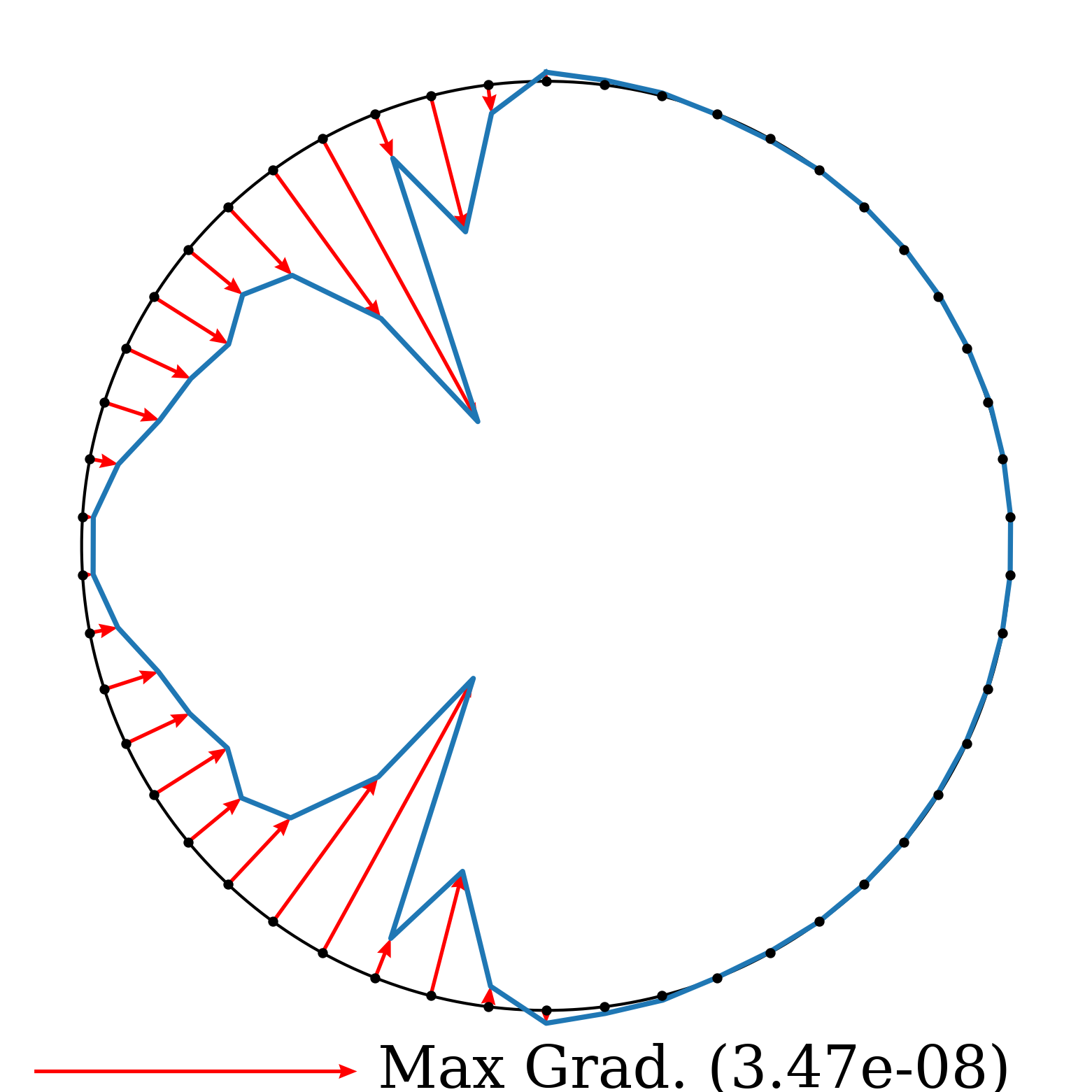}}
    \caption{Adjoint HRRLBM-LES model with $\tau_{\mathrm{adj}} = 0.65$.}
    \label{plot:surfaceARLBMtau0.65}
    \end{subfigure}
    
    \begin{subfigure}[b]{.49\textwidth}
    \centering
    \raisebox{0.mm}{\includegraphics[width=0.6\linewidth]{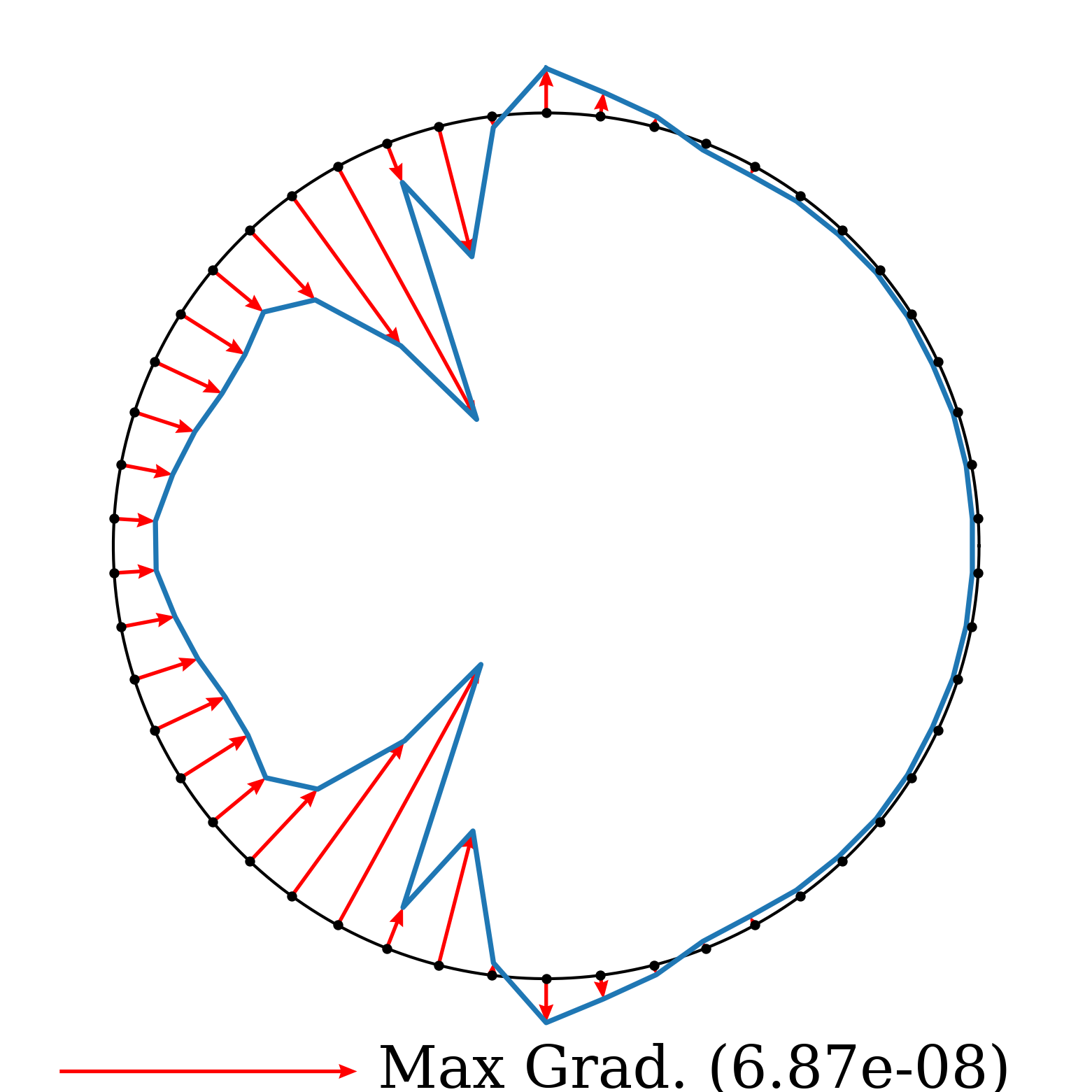}}
    \caption{Adjoint HRRLBM model with FTA with $\tau_{\mathrm{adj}} = 0.6$.}
    \label{plot:surfaceFTAtau0.6}
    \end{subfigure}
    \hfill
    \begin{subfigure}[b]{.49\textwidth}
    \centering
    \raisebox{0.mm}{\includegraphics[width=0.6\linewidth]{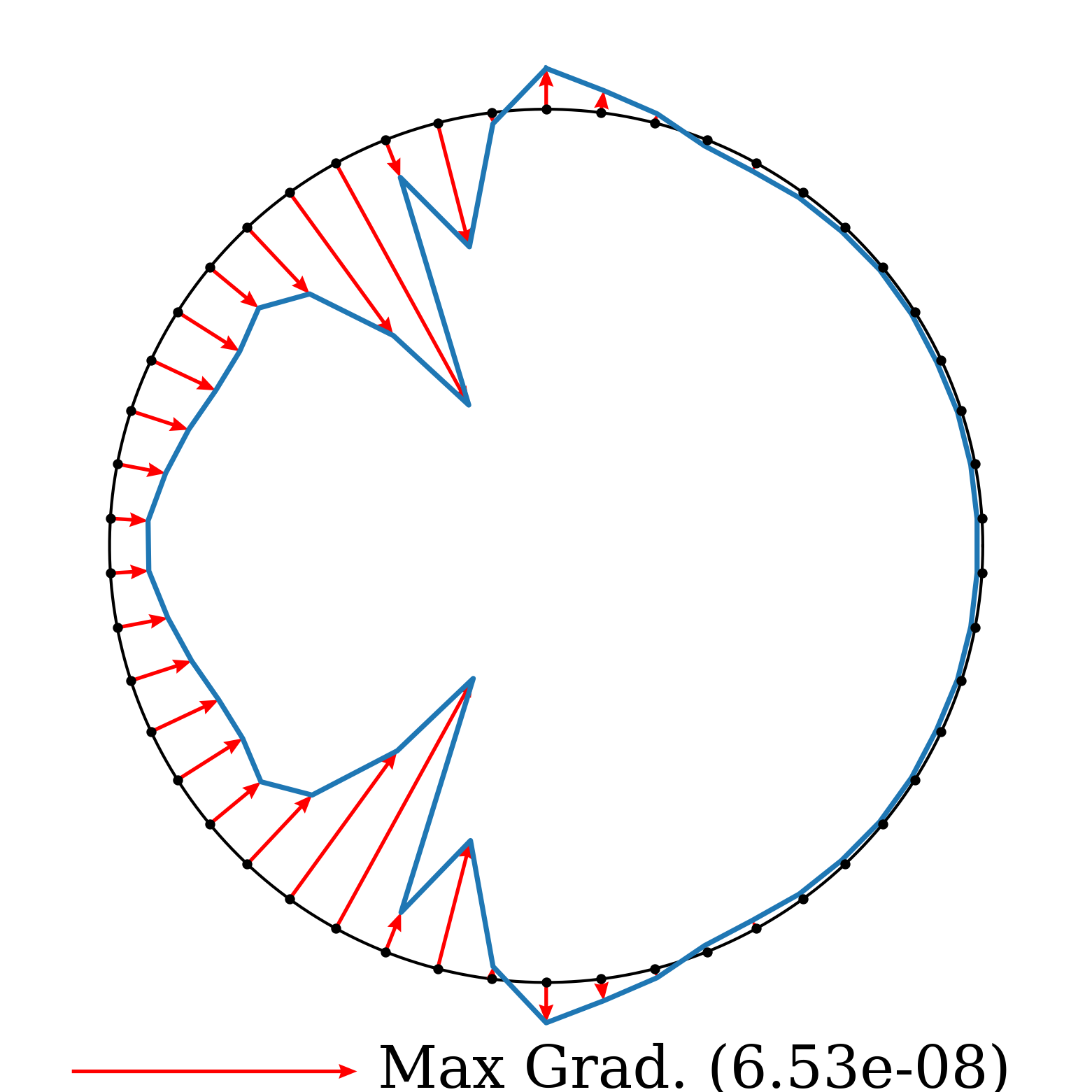}}
    \caption{Adjoint HRRLBM model with FTA with $\tau_{\mathrm{adj}} = 0.65$.}
    \label{plot:surfaceFTAtau0.65}
    \end{subfigure}
    \caption{Surface gradients computed by the mean-flow adjoint approach for the turbulent cylinder case at $Re = 3900$ projected onto the surface normals.}
    \label{plot:surface}
\end{figure}

The adjoint surface gradients are projected onto the normal vectors of the cylinder in Fig.~\ref{plot:surface}.
In Fig.~\ref{plot:surfaceARLBMtau0.6} and Fig.~\ref{plot:surfaceARLBMtau0.65}, the gradients are visualized for the completely differentiated primal model for the adjoint lattice relaxation times of $\tau_{\mathrm{adj}} = 0.6$ and $\tau_{\mathrm{adj}} = 0.65$.
Analogously, in Fig.~\ref{plot:surfaceFTAtau0.6} and Fig.~\ref{plot:surfaceFTAtau0.65}, the gradients are visualized for the adjoint simulation with the FTA again for the adjoint lattice relaxation times of $\tau_{\mathrm{adj}} = 0.6$ and $\tau_{\mathrm{adj}} = 0.65$.
The adjoint simulation with the FTA at $\tau_{\mathrm{adj}} = 0.55$ still produced smooth surface gradients, while for the fully differentiated case, the adjoint simulation diverged.
This matches the literature reports, stating that the fully differentiated non-linear LES models may raise stability concerns in the adjoint simulations~\cite{MARTA2013102, blonigan2012towards}.
Looking at the HRRLBM-LES equation in~\eqref{eq:HRRLBM} and the adjoint update scheme in~\eqref{eq:adjoint_unsteady}, the adjoint LBM is obtained by differentiating the primal model by the particle distribution function.
Due to the present LES model, the effective relaxation time is dependent on $f$ such that computing $\partial \mathcal{C}/\partial f$ gives

\begin{align}
    \frac{\partial \mathcal{C}_j}{\partial f_i} = \frac{\partial f_j^{\mathrm{eq}}}{\partial f_i} + \left(1-\frac{1}{\tau_{\mathrm{eff}}}\right)\frac{\partial f_j^{(1)}}{\partial f_i} + \left(\frac{1}{\tau_{\mathrm{eff}}}\right)^2\frac{\partial \tau_{\mathrm{eff}}}{\partial f_i}f_j^{(1)}.
\end{align}

\noindent The last term on the right-hand side is an additional source term, emerging in the adjoint LBM due to the present LES model.
It is proportional to the non-equilibrium parts, which become more dominant in regions with high shear rates in the primal problem, such as in the vicinity of the cylinder surface.
Therefore, depending on the LES model, this term possibly inserts energy in the high-shear regions of the flow, destabilizing the adjoint simulation.
This matches our observations, as in adjoint simulations, where stability issues occurred, the source of spurious oscillations was the close vicinity of the cylinder surface.
This example shows that a LES model, which generally exhibits stabilizing effects in the primal model, does not necessarily stabilize the adjoint problem.
Comparing Fig.~\ref{plot:surfaceARLBMtau0.6} and Fig.~\ref{plot:surfaceARLBMtau0.65}, as well as Fig.~\ref{plot:surfaceFTAtau0.6} and Fig.~\ref{plot:surfaceFTAtau0.65}, shows that the regularization mainly influences the magnitude of the surface gradients while the distribution on the surface remains similar.
In all cases displayed in Fig.~\ref{plot:surface}, the surface gradients suggest changes in the cylinder topology such that the upstream face becomes more aligned to the fluid flow to reduce the drag. 
Comparing the results for the fully differentiated LES model and the FTA, differences in the surface gradients regarding the magnitude as well as the distribution are observable.
At the top and bottom side of the cylinder, the FTA gradients suggest to increase the hydraulic diameter by adding material outwards, which is not visible for the gradients obtained by the fully differentiated LES model.
This corresponds to the observations made for RANS-based adjoint sensitivity analysis, that the FTA might lead to strong deviations in the computed gradients as demonstrated in~\cite{computation6010005}.
As the comparison of the surface gradients for this case with FDQ or forward AD is not reasonable, a feasibility study of those sensitivities performing topology optimization is the target of future work.
\section{Conclusion\label{sec:conclusion}}

In this work, an adjoint sensitivity analysis was conducted for steady and unsteady flow around porous cylinders modeled via the HLBM. 
The first-discretize-then-differentiate approach utilizes AD for the Jacobian expressions in the adjoint problem and was successfully evaluated across different flow regimes.
The validation studies in the steady laminar regime at $Re=20$ (2D) and $Re=40$ (3D) demonstrated that the computed adjoint sensitivities match the gradients obtained from FDQ and forward AD with high accuracy.
Reducing the floating-point precision in the 3D cases showed no visible difference in the validation results, while drastically reducing the numerical cost.
Regarding the investigated objective functionals, the drag evaluation based on the FFA and the energy dissipation functional yielded robust and accurate gradients.
In contrast, the drag evaluation using the MEA suffered from instabilities and gradient deviations close to the cylinder wall, which aligns with literature recommendations to separate the control and objective domains from the solid body boundaries.
For the unsteady laminar regime at $Re=100$, where the flow develops a transient von Kármán vortex street, the mean-flow adjoint LBM approach proposed by Cheylan et al. was successfully applied.
By utilizing the time-averaged primal particle distribution functions to evaluate the local evolution operators, the framework circumvented the massive memory footprint of exact transient checkpointing.
The resulting mean-flow adjoint velocity distributions and surface sensitivities showed symmetrical profiles and matched the structural observations reported in prior mean-flow literature very well.
This demonstrates the applicability of combining mean-flow adjoints with the Brinkman-penalization-based porous cylinder model.
Finally, a proof-of-concept study was performed for a chaotic, turbulent free flow around a 3D cylinder at $Re=3900$ utilizing the HRRLBM-LES model.
To counteract the inherent trajectory divergence caused by positive Lyapunov exponents in chaotic regimes, the mean-flow-based adjoint gradient computation is applied.
A comparison between the completely differentiated LES operator and the FTA revealed that the FTA requires less artificial numerical diffusion to stabilize the adjoint simulation.
An additional source term emerges due to the LES model being dependent on the state variable, where we observed spurious oscillations appeared around the cylinder in case the adjoint simulation was unstable.
Regularization measures by adding numerical diffusion are applied to stabilize the adjoint simulation by increasing the adjoint lattice relaxation time ($\tau_{adj} = 0.6$ and $0.65$).
They mainly scaled the magnitude of the surface gradients while the change in gradient distribution was small.
A definitive feasibility study of these turbulent sensitivities should be conducted in future work by implementing the computed gradients directly into a gradient-based optimizer to perform fluid topology optimization.

\section*{Author contributions statement}
Conceptualization: SI;
Methodology: SI, FB;
Software: SI, AK;
Validation: SI;
Formal analysis: SI;
Investigation: SI;
Resources: MJK;
Data Curation: SI;
Writing - Original Draft: SI;
Writing - Review \& Editing: SI, JLG;
Visualization: SI;
Supervision: MJK;
Project administration: MJK;
Funding acquisition: MJK.
All authors read and approved the final version of the paper.

\section*{Funding}
This work has received funding from the European Union’s Horizon Europe research and innovation program under grant agreement No 101138305. % FALCON

\section*{Acknowledgments}
This work was performed on the HoreKa supercomputer funded by the
Ministry of Science, Research and the Arts Baden-Württemberg and by
the Federal Ministry of Education and Research.
The authors thank Nicolas R. Gauger and Max Sagebaum from the RPTU University Kaiserslautern-Landau and Isabelle Cheylan from the Aix-Marseille University for fruitful discussions on the adjoint-based sensitivity computation under turbulence.

\section*{Competing interests statement} 
The authors declare that they have no known competing financial interests or personal relationships that could have appeared to influence the work reported in this paper.

\section*{Data availability}
The results in this paper were produced with \textit{OpenLB} \cite{krause_openlbopen_2021} (published under GNU General Public License V.2 (GPL2), version 1.9~\cite{kummerlander_2025_17899765}).
The computational data are available upon reasonable request.

\appendix
\section{Temporal average of objective functionals and collision operators}\label{app:tempAvg}

Let $\mathcal{F}(\bm{\alpha}, f)$ an arbitrary function dependent on a time-invariant control vector $\bm{\alpha}$ and time dependent state function $f$.
The state function is split into temporal average and fluctuations as $f = f^{'} +\overline{f}$ with $\overline{f^{'}} = 0$.
The Taylor-series expansion of the function $\mathcal{F}$ around the time-invariant state gives:

\begin{align}
    \mathcal{F}(\bm{\alpha}, f) = \mathcal{F}(\bm{\alpha}, \overline{f}) + \left. \frac{\partial \mathcal{F}(\bm{\alpha}, f)}{\partial f} \right|_{f = \overline{f}}f^{'} + \frac{1}{2}\left. \frac{\partial^2 \mathcal{F}(\bm{\alpha}, f)}{\partial f^2} \right|_{f = \overline{f}}(f^{'})^2 + \mathcal{O}((f^{'})^3).
\end{align}

\noindent Applying temporal averaging over the series expansion, it yields

\begin{align}
    \overline{\mathcal{F}(\bm{\alpha}, f)} = \mathcal{F}(\bm{\alpha}, \overline{f}) + \frac{1}{2}\left. \frac{\partial^2 \mathcal{F}(\bm{\alpha}, f)}{\partial f^2} \right|_{f = \overline{f}}\overline{(f^{'})^2} + \mathcal{O}((f^{'})^3).
\end{align}

\noindent This shows, that if $\mathcal{F}$ is linear regarding $f$, the equality $\overline{\mathcal{F}(\bm{\alpha}, f)} = \mathcal{F}(\bm{\alpha}, \overline{f})$ holds.
Replacing $\mathcal{F}$ with $\partial J/\partial f$, $\partial J/\partial \bm{\alpha}$, $\partial \mathcal{C}/\partial f$, and $\partial \mathcal{C}/\partial \bm{\alpha}$ gives the explanation of the statement given in Sec.~\ref{sec:unsteady_opti}.

\section{Influence of the floating point precision}\label{app:FPP}

\begin{figure}[h!]
    \centering
    \includegraphics[width=0.8\linewidth]{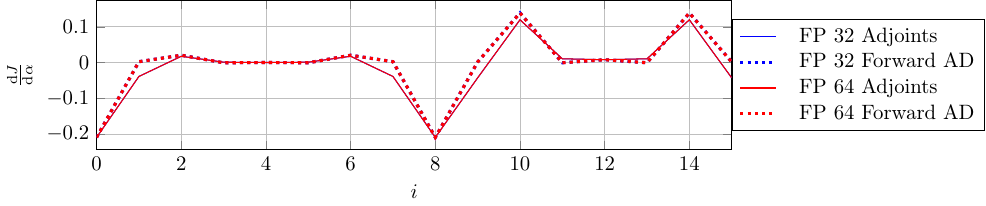}
    \caption{Comparison of adjoint gradients with forward AD for different floating-point precision for the case~\ref{sec:adjointValidation3D}. FDQ yielded inaccurate results due to too large stencil widths.}
    \label{fig:FPP_comp}
\end{figure}

To assess the impact of the floating point precision, the validation case in Sec.~\ref{sec:adjointValidation3D} is computed for single and double precision.
Fig.~\ref{fig:FPP_comp} demonstrates that for both the adjoint and forward AD-based gradients, the results match perfectly for the different floating point precisions.

\section{Divergence of exact transient sensitivities via forward AD}
\label{appendix:forward_ad_divergence}

To further justify the necessity of the mean-flow adjoint approach for unsteady configurations, this appendix analyzes the behavior of exact transient sensitivities computed via forward AD for the case from Sec.~\ref{sec:meanFlowAdjoints} at $Re = 100$. 
When evaluating the exact transient sensitivity of a time-integrated objective functional (such as drag) over a simulation time horizon $T$, a well-known phenomenon of gradient explosion as in~\cite{LES2016} is observed.

\begin{figure}[h!]
    \centering
    \begin{subfigure}[t]{.49\textwidth}
    \raisebox{0.mm}{\includegraphics[width=1.\linewidth]{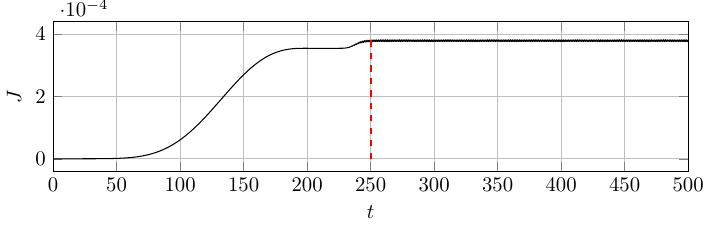}}
    \caption{Dissipation objective functional over simulation time. The dashed red line indicates the start of the temporal averaging for the tracking average type objective functional.}
    \label{plot:objectiveRe100}
    \end{subfigure}
    \hfill
    \begin{subfigure}[t]{.49\textwidth}
    \raisebox{1.5mm}{\includegraphics[width=1.\linewidth]{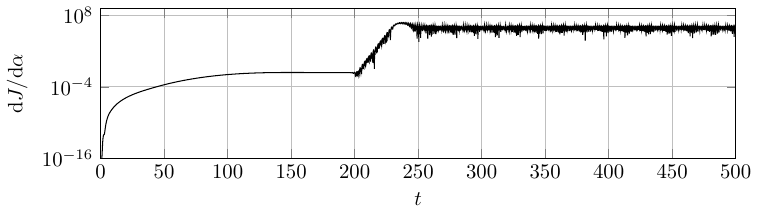}}
    \caption{Transient development of the forward AD gradients, where the initial seeding of the AD algorithm is performed at $t = 0$.}
    \label{plot:gradRe100}
    \end{subfigure}
    \caption{Exploding forward AD sensitivities for the 2D cylinder flow at $Re = 100$.}
    \label{plot:forwardADRe100}
\end{figure}

\noindent Fig.~\ref{plot:objectiveRe100} shows the temporal evolution of the dissipation objective over time from~\eqref{eq:obj_diss} by evaluating the function for the state at every time step $t$.
Therein, between $t \in [200, 250]$s, the Helmholtz instabilities trigger the periodic vortex street for $t \geq 250$s.
In the same time span, the forward AD sensitivities mark an increase in the gradients of the magnitude of $10^9$, as shown in Fig.~\ref{plot:gradRe100}.
To verify this observation, the wake velocity component perpendicular to the main flow direction $u_y$ of two successive simulations with perturbation applied for the controlled permeability of a single cell on the cylinder surface with $\triangle \alpha = 1e-6$ m$^2$ are compared.

\begin{figure}[h!]
    \centering
    \includegraphics[width=0.6\linewidth]{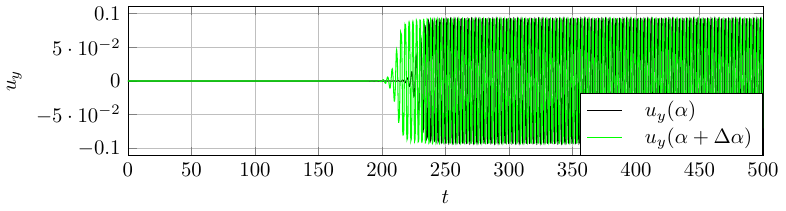}
    \caption{Comparison of the wake velocity y-component for slightly perturbed cylinder topologies to demonstrate the shift in the onset of the Helmholtz instabilities.}
    \label{plot:shift}
\end{figure}

\noindent Fig.~\ref{plot:shift} illustrates that slight changes in the local permeability lead to significant temporal shifts in the onset of the periodic vortex street, which directly lead to the instabilities of the forward AD algorithm gradients. 

The same investigations are performed for the case, where the seeding of the forward AD algorithm takes place after the full development of the periodic vortex shredding, i.e., at $t = 250$s.
This means that the forward AD algorithm skips the critical part between $t \in [200, 250]$s.

\begin{figure}[h!]
    \centering
    \begin{subfigure}[t]{.7\textwidth}
    \raisebox{0.mm}{\includegraphics[width=1.\linewidth]{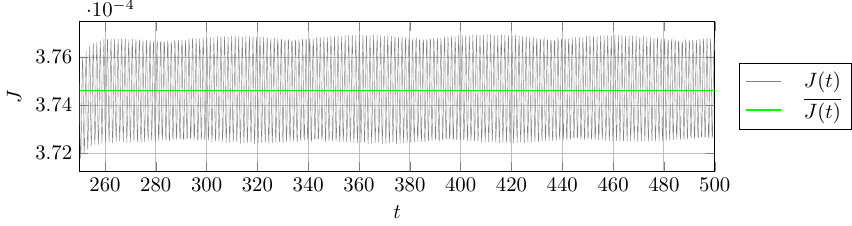}}
    \caption{Dissipation objective functional over simulation time after $t = 250$s. The solid green line shows the temporal average of the objective functional.}
    \label{plot:averageObj}
    \end{subfigure}
    \vfill
    \begin{subfigure}[t]{.7\textwidth}
    \raisebox{0mm}{\includegraphics[width=1.\linewidth]{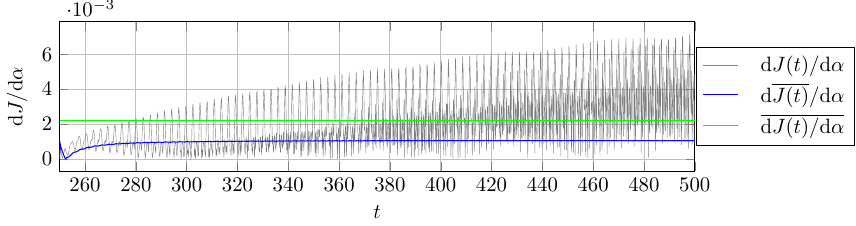}}
    \caption{Transient development of the forward AD gradients, where the initial seeding of the AD algorithm is performed after the full development of the periodic vortex shredding (at $t = 250$s). The blue line gives the objective derivative of the tracking-type temporal average, whereas the green line gives the average of the objective sensitivities as plotted in black.}
    \label{plot:phaseShift}
    \end{subfigure}
    \caption{Exploding forward AD sensitivities for the 2D cylinder flow at $Re = 100$.}
    \label{plot:phaseShiftRe100}
\end{figure}

Fig.~\ref{plot:averageObj} visualizes the objective functional and the temporal average of the objective, where the averaged value matches the mean value of the oscillating fluctuations.
However, in Fig.~\ref{plot:phaseShift}, we observe that the forward AD sensitivities slowly increase over time.
This is confirmed by comparing the derivative of the averaged objective functional computed by \eqref{eq:averageObjective} and the average of the derivatives plotted by the black solid line in the current figure.
We observe here the exact behavior reported by~\cite{LSS2018} that

\begin{align}
    \frac{\partial}{\partial \bm{\alpha}}\left(\overline{J(\bm{\alpha}, f(\cdot, t))}\right) \neq \overline{\left(\frac{\partial J(\bm{\alpha}, f(\cdot, t))}{\partial \bm{\alpha}}\right)},
\end{align}

\noindent even for this laminar unsteady case for the cylinder flow at $Re = 100$.
A minute change in frequency leads to perturbed and unperturbed flow trajectories that become increasingly out of phase as time progresses.
The forward AD gradients track these diverging trajectories by experiencing oscillatory growth where the magnitude increases monotonically with the time horizon $T$.

Summarized, the divergence of the forward AD gradients is primarily triggered by two coupled mechanisms:
\begin{enumerate}
    \item Minor variations in the local lattice porosity or solid boundary definitions induce a temporal shift in the physical onset of the Helmholtz instabilities behind the cylinder. 
    \item The design variables inherently alter the shedding frequency (Strouhal number) and the phase of the transient vortex street, leading to accumulating contributions for the objective sensitivities.
\end{enumerate}

\bibliographystyle{elsarticle-num}
\bibliography{paper}

@article{mohammadi2004shape,
  title={Shape optimization in fluid mechanics},
  author={Mohammadi, Bijan and Pironneau, Olivier},
  journal={Annu. Rev. Fluid Mech.},
  volume={36},
  number={1},
  pages={255--279},
  year={2004},
  publisher={Annual Reviews}
}

@Article{computation6010005,
AUTHOR = {Schramm, Matthias and Stoevesandt, Bernhard and Peinke, Joachim},
TITLE = {Optimization of Airfoils Using the Adjoint Approach and the Influence of Adjoint Turbulent Viscosity},
JOURNAL = {Computation},
VOLUME = {6},
YEAR = {2018},
NUMBER = {1},
ARTICLE-NUMBER = {5},
URL = {https://www.mdpi.com/2079-3197/6/1/5},
ISSN = {2079-3197},
DOI = {10.3390/computation6010005}
}

@article{Qian.1992,
 author = {Qian, Y. H. and D'Humi{\`e}res, D. and Lallemand, P.},
 year = {1992},
 title = {Lattice BGK Models for Navier-Stokes Equation},
 pages = {479--484},
 volume = {17},
 number = {6},
 issn = {0295-5075},
 journal = {Europhysics Letters (EPL)},
 doi = {10.1209/0295-5075/17/6/001}
}

@article{Bhatnagar.1954,
 author = {Bhatnagar, P. L. and Gross, E. P. and Krook, M.},
 year = {1954},
 title = {A Model for Collision Processes in Gases. I. Small Amplitude Processes in Charged and Neutral One-Component Systems},
 pages = {511--525},
 volume = {94},
 number = {3},
 issn = {0031-899X},
 journal = {Physical Review},
 doi = {10.1103/PhysRev.94.511}
}

@article{coreixas2017recursive,
  title={Recursive regularization step for high-order lattice Boltzmann methods},
  author={Coreixas, Christophe and Wissocq, Gauthier and Puigt, Guillaume and Boussuge, Jean-Fran{\c{c}}ois and Sagaut, Pierre},
  journal={arXiv preprint arXiv:1704.04413},
  year={2017}
}

@article{Wang_Gao_2013, title={The drag-adjoint field of a circular cylinder wake at Reynolds numbers 20, 100 and 500}, volume={730}, DOI={10.1017/jfm.2013.323}, journal={Journal of Fluid Mechanics}, author={Wang, Qiqi and Gao, Jun-Hui}, year={2013}, pages={145–161}}

@software{kummerlander_2025_17899765,
  author       = {Kummerländer, Adrian and
                  Bingert, Tim and
                  Bock, Simon and
                  Bukreev, Fedor and
                  Castroviejo, Daniel and
                  Czelusniak, Luiz Eduardo and
                  Dapelo, Davide and
                  Gaul, Christoph and
                  Dorn, Marcio and
                  Dorneles, Leonardo and
                  Grafen, Johannes and
                  Grinschewski, Michael and
                  Ito, Shota and
                  Jeßberger, Julius and
                  Kaiser, Florian and
                  Khazaeipoul, Danial and
                  Krüger, Timm and
                  Kumbhat, Arsh and
                  Kusumaatmaja, Halim and
                  Nettekoven, Andreas and
                  Raeli, Alice and
                  Riazantsev, Tikhon and
                  Rennick, Michael and
                  Prakash, Gagan and
                  Prinz, František and
                  Sauterleute, Liam and
                  Schecher, Maximilian and
                  Schneider, Andreas and
                  Shimojima, Yuji and
                  Simonis, Stephan and
                  Spelten, Philipp and
                  Tacques, Alexandre and
                  Krause, Mathias J.},
  title        = {OpenLB Release 1.9: Open Source Lattice Boltzmann
                   Code
                  },
  month        = dec,
  year         = 2025,
  publisher    = {Zenodo},
  version      = {1.9.0},
  doi          = {10.5281/zenodo.17899765},
  url          = {https://doi.org/10.5281/zenodo.17899765},
}

@inproceedings{blonigan2012towards,
  title={Towards adjoint sensitivity analysis of statistics in turbulent flow simulation},
  author={Blonigan, P and Chen, R and Wang, Q and Larsson, J},
  booktitle={Proceedings of the Summer Program},
  volume={229},
  year={2012},
  organization={Center for Turbulence Research, Stanford Univ.}
}

@article{FRANKE20021191,
title = {Large eddy simulation of the flow past a circular cylinder at ReD=3900},
journal = {Journal of Wind Engineering and Industrial Aerodynamics},
volume = {90},
number = {10},
pages = {1191-1206},
year = {2002},
note = {3rd European-African Conference on Wind Engineering},
issn = {0167-6105},
doi = {https://doi.org/10.1016/S0167-6105(02)00232-5},
url = {https://www.sciencedirect.com/science/article/pii/S0167610502002325},
author = {J. Franke and W. Frank}
}

@article {LES2016,
author = {Rajani, B. N. and Kandasamy, A. and Majumdar, Sekhar},
title = {LES of Flow past Circular Cylinder at Re = 3900},
journal = {Journal of Applied Fluid Mechanics},
volume = {9},
number = {3},
pages = {1421-1435},
year  = {2016},
publisher = {},
issn = {1735-3572}, 
eissn = {1735-3645}, 
doi = {10.18869/acadpub.jafm.68.228.24178},	
url = {https://www.jafmonline.net/article_1718.html},
eprint = {https://www.jafmonline.net/article_1718_fc6709bfdf0572f183c1a84ce5276e96.pdf}
}

@article{abernathy1962formation,
  title={The formation of vortex streets},
  author={Abernathy, Frederick H and Kronauer, Richard E},
  journal={Journal of Fluid Mechanics},
  volume={13},
  number={1},
  pages={1--20},
  year={1962},
  publisher={Cambridge University Press}
}

@book{nadarajah2003discrete,
  title={The discrete adjoint approach to aerodynamic shape optimization},
  author={Nadarajah, Siva Kumaran},
  year={2003},
  publisher={stanford university}
}

@article{bouzidi2001momentum,
  title={Momentum transfer of a Boltzmann-lattice fluid with boundaries},
  author={Bouzidi, M’hamed and Firdaouss, Mouaouia and Lallemand, Pierre},
  journal={Physics of fluids},
  volume={13},
  number={11},
  pages={3452--3459},
  year={2001},
  publisher={American Institute of Physics}
}

@incollection{schafer1996benchmark,
  title={Benchmark computations of laminar flow around a cylinder},
  author={Sch{\"a}fer, Michael and Turek, Stefan and Durst, Franz and Krause, Egon and Rannacher, Rolf},
  booktitle={Flow simulation with high-performance computers II: DFG priority research programme results 1993--1995},
  pages={547--566},
  year={1996},
  publisher={Springer}
}

@article{Wang2011,
author = {Wang, Chen-Hao and Ho, Jeng-Rong},
year = {2011},
month = {07},
pages = {75-86},
title = {A lattice Boltzmann approach for the non-Newtonian effect in the blood flow},
volume = {62},
journal = {Computers \& Mathematics with Applications},
doi = {10.1016/j.camwa.2011.04.051}
}

@article{laddNumericalSimulationsParticulate1994,
  title = {Numerical Simulations of Particulate Suspensions via a Discretized {{Boltzmann}} Equation. {{Part}} 1. {{Theoretical}} Foundation},
  author = {Ladd, Anthony J. C.},
  date = {1994-07-25},
  journaltitle = {Journal of Fluid Mechanics},
  shortjournal = {J. Fluid Mech.},
  volume = {271},
  pages = {285--309},
  issn = {0022-1120, 1469-7645},
  doi = {10.1017/S0022112094001771},
  url = {https://www.cambridge.org/core/product/identifier/S0022112094001771/type/journal_article},
  urldate = {2025-02-13},
  langid = {english}
}

@article{laddNumericalSimulationsParticulate1994a,
  title = {Numerical Simulations of Particulate Suspensions via a Discretized {{Boltzmann}} Equation. {{Part}} 2. {{Numerical}} Results},
  author = {Ladd, Anthony J. C.},
  date = {1994-07-25},
  journaltitle = {Journal of Fluid Mechanics},
  shortjournal = {J. Fluid Mech.},
  volume = {271},
  pages = {311--339},
  issn = {0022-1120, 1469-7645},
  doi = {10.1017/S0022112094001783},
  url = {https://www.cambridge.org/core/product/identifier/S0022112094001783/type/journal_article},
  urldate = {2025-02-06},
  langid = {english}
}

@article{ONORATO1986317,
title = {Experimental analysis of vehicle wakes},
journal = {Journal of Wind Engineering and Industrial Aerodynamics},
volume = {22},
number = {2},
pages = {317-330},
year = {1986},
note = {Special Issue 6th Colloquium on Industrial Aerodynamics Vehicle Aerodynamics},
issn = {0167-6105},
doi = {https://doi.org/10.1016/0167-6105(86)90094-2},
url = {https://www.sciencedirect.com/science/article/pii/0167610586900942},
author = {M. Onorato and A.F. Costelli and A. Garrone and L. Viassone}
}

@book{gunzburger2002,
author = {Gunzburger, Max D.},
title = {Perspectives in Flow Control and Optimization},
publisher = {Society for Industrial and Applied Mathematics},
year = {2002},
doi = {10.1137/1.9780898718720},
address = {},
edition   = {},
URL = {https://epubs.siam.org/doi/abs/10.1137/1.9780898718720},
eprint = {https://epubs.siam.org/doi/pdf/10.1137/1.9780898718720}
}

@article{smagorinsky1963general,
  title={General circulation experiments with the primitive equations: I. The basic experiment},
  author={Smagorinsky, Joseph},
  journal={Monthly weather review},
  volume={91},
  number={3},
  pages={99--164},
  year={1963}
}

@article{Malaspinas_Sagaut_2012, title={Consistent subgrid scale modelling for lattice Boltzmann methods}, volume={700}, DOI={10.1017/jfm.2012.155}, journal={Journal of Fluid Mechanics}, author={Malaspinas, Orestis and Sagaut, Pierre}, year={2012}, pages={514–542}}

@article{jacob2018new,
  title={A new hybrid recursive regularised Bhatnagar--Gross--Krook collision model for lattice Boltzmann method-based large eddy simulation},
  author={Jacob, J{\'e}r{\^o}me and Malaspinas, Orestis and Sagaut, Pierre},
  journal={Journal of Turbulence},
  volume={19},
  number={11-12},
  pages={1051--1076},
  year={2018},
  publisher={Taylor \& Francis}
}

@article{spaid1997lattice,
  title={Lattice Boltzmann methods for modeling microscale flow in fibrous porous media},
  author={Spaid, Michael AA and Phelan Jr, Frederick R},
  journal={Physics of fluids},
  volume={9},
  number={9},
  pages={2468--2474},
  year={1997},
  publisher={American Institute of Physics}
}

@book{Kruger.2017,
 author = {Kr{\"u}ger, Timm and Kusumaatmaja, Halim and Kuzmin, Alexandr and Shardt, Orest and Silva, Goncalo and Viggen, Erlend Magnus},
 year = {2017},
 title = {The Lattice Boltzmann Method: Principles and Practice},
 address = {Cham},
 publisher = {{Springer International Publishing} and Imprint and Springer},
 isbn = {9783319446493},
 series = {Graduate Texts in Physics}
}

@article{simonis2025homogenized,
  title={Homogenized lattice Boltzmann methods for fluid flow through porous media--part I: kinetic model derivation},
  author={Simonis, Stephan and Hafen, Nicolas and Je{\ss}berger, Julius and Dapelo, Davide and Th{\"a}ter, Gudrun and Krause, Mathias J},
  journal={ESAIM: Mathematical Modelling and Numerical Analysis},
  volume={59},
  number={2},
  pages={789--813},
  year={2025},
  publisher={EDP Sciences}
}

@book{stuck2012adjoint,
  title={Adjoint Navier-Stokes methods for hydrodynamic shape optimisation},
  author={St{\"u}ck, Arthur},
  year={2012},
  publisher={Technische Universit{\"a}t Hamburg}
}

@article{dwight2006,
author = {Dwight, Richard P. and Brezillon, Joel},
title = {Effect of Approximations of the Discrete Adjoint on Gradient-Based Optimization},
journal = {AIAA Journal},
volume = {44},
number = {12},
pages = {3022-3031},
year = {2006},
doi = {10.2514/1.21744},

URL = { 
    
        https://doi.org/10.2514/1.21744
    
    

},
eprint = { 
    
        https://doi.org/10.2514/1.21744
    
    

}
}

@article{MARTA2013102,
title = {On the handling of turbulence equations in RANS adjoint solvers},
journal = {Computers \& Fluids},
volume = {74},
pages = {102-113},
year = {2013},
issn = {0045-7930},
doi = {https://doi.org/10.1016/j.compfluid.2013.01.012},
url = {https://www.sciencedirect.com/science/article/pii/S0045793013000303},
author = {Andre C. Marta and Sriram Shankaran}
}

@article{papoutsis2016continuous,
  title={Continuous adjoint methods for turbulent flows, applied to shape and topology optimization: industrial applications},
  author={Papoutsis-Kiachagias, Evangelos M and Giannakoglou, Kyriakos C},
  journal={Archives of Computational Methods in Engineering},
  volume={23},
  number={2},
  pages={255--299},
  year={2016},
  publisher={Springer}
}

@article{LSS2018,
author = {Blonigan, Patrick J. and Wang, Qiqi and Nielsen, Eric J. and Diskin, Boris},
title = {Least-Squares Shadowing Sensitivity Analysis of Chaotic Flow Around a Two-Dimensional Airfoil},
journal = {AIAA Journal},
volume = {56},
number = {2},
pages = {658-672},
year = {2018},
doi = {10.2514/1.J055389},

URL = { 
    
        https://doi.org/10.2514/1.J055389
    
    

},
eprint = { 
    
        https://doi.org/10.2514/1.J055389
    
    

}
}

@article{meliga2014,
    author = {Meliga, Philippe and Boujo, Edouard and Pujals, Gregory and Gallaire, François},
    title = {Sensitivity of aerodynamic forces in laminar and turbulent flow past a square cylinder},
    journal = {Physics of Fluids},
    volume = {26},
    number = {10},
    pages = {104101},
    year = {2014},
    month = {10},
    issn = {1070-6631},
    doi = {10.1063/1.4896941},
    url = {https://doi.org/10.1063/1.4896941},
    eprint = {https://pubs.aip.org/aip/pof/article-pdf/doi/10.1063/1.4896941/13799256/104101_1_online.pdf},
}

@article{yaji2018large,
  title={Large-scale topology optimization incorporating local-in-time adjoint-based method for unsteady thermal-fluid problem},
  author={Yaji, Kentaro and Ogino, Masao and Chen, Cong and Fujita, Kikuo},
  journal={Structural and Multidisciplinary Optimization},
  volume={58},
  number={2},
  pages={817--822},
  year={2018},
  publisher={Springer}
}

@article{YAMALEEV20105394,
title = {Local-in-time adjoint-based method for design optimization of unsteady flows},
journal = {Journal of Computational Physics},
volume = {229},
number = {14},
pages = {5394-5407},
year = {2010},
issn = {0021-9991},
doi = {https://doi.org/10.1016/j.jcp.2010.03.045},
url = {https://www.sciencedirect.com/science/article/pii/S0021999110001646},
author = {Nail K. Yamaleev and Boris Diskin and Eric J. Nielsen}
}

@article{NORGAARD2016291,
title = {Topology optimization of unsteady flow problems using the lattice Boltzmann method},
journal = {Journal of Computational Physics},
volume = {307},
pages = {291-307},
year = {2016},
issn = {0021-9991},
doi = {https://doi.org/10.1016/j.jcp.2015.12.023},
url = {https://www.sciencedirect.com/science/article/pii/S0021999115008426},
author = {Sebastian Nørgaard and Ole Sigmund and Boyan Lazarov}
}

@article{KRAUSE20171,
title = {Particle flow simulations with homogenised lattice Boltzmann methods},
journal = {Particuology},
volume = {34},
pages = {1-13},
year = {2017},
issn = {1674-2001},
doi = {https://doi.org/10.1016/j.partic.2016.11.001},
url = {https://www.sciencedirect.com/science/article/pii/S167420011730041X},
author = {Mathias J. Krause and Fabian Klemens and Thomas Henn and Robin Trunk and Hermann Nirschl}
}

@article{borrvall2003,
author = {Borrvall, Thomas and Petersson, Joakim},
title = {Topology optimization of fluids in Stokes flow},
journal = {International Journal for Numerical Methods in Fluids},
volume = {41},
number = {1},
pages = {77-107},
doi = {https://doi.org/10.1002/fld.426},
url = {https://onlinelibrary.wiley.com/doi/abs/10.1002/fld.426},
eprint = {https://onlinelibrary.wiley.com/doi/pdf/10.1002/fld.426},
year = {2003}
}

@article{YAJI2014158,
title = {Topology optimization using the lattice Boltzmann method incorporating level set boundary expressions},
journal = {Journal of Computational Physics},
volume = {274},
pages = {158-181},
year = {2014},
issn = {0021-9991},
doi = {https://doi.org/10.1016/j.jcp.2014.06.004},
url = {https://www.sciencedirect.com/science/article/pii/S0021999114004112},
author = {Kentaro Yaji and Takayuki Yamada and Masato Yoshino and Toshiro Matsumoto and Kazuhiro Izui and Shinji Nishiwaki}
}

@article{cheylan_shape_2019,
	title = {Shape {Optimization} {Using} the {Adjoint} {Lattice} {Boltzmann} {Method} for {Aerodynamic} {Applications}},
	volume = {57},
	issn = {0001-1452},
	url = {https://doi.org/10.2514/1.J057955},
	doi = {10.2514/1.J057955},
	number = {7},
	urldate = {2026-06-24},
	journal = {AIAA Journal},
	author = {Cheylan, Isabelle and Fritz, Guillaume and Ricot, Denis and Sagaut, Pierre},
	year = {2019},
	note = {Publisher: American Institute of Aeronautics and Astronautics
\_eprint: https://doi.org/10.2514/1.J057955},
	pages = {2758--2773},
}

@article{jalali_khouzani_airfoil_2023,
	title = {Airfoil inverse design based on laminar compressible adjoint lattice {Boltzmann} method},
	volume = {95},
	copyright = {© 2023 John Wiley \& Sons Ltd.},
	issn = {1097-0363},
	url = {https://onlinelibrary.wiley.com/doi/abs/10.1002/fld.5192},
	doi = {10.1002/fld.5192},
	language = {en},
	number = {8},
	urldate = {2026-06-24},
	journal = {International Journal for Numerical Methods in Fluids},
	author = {Jalali Khouzani, H. and Kamali-Moghadam, R.},
	year = {2023},
	note = {\_eprint: https://onlinelibrary.wiley.com/doi/pdf/10.1002/fld.5192},
	pages = {1197--1219},
}

@article{chen_local--time_2017,
	title = {Local-in-time adjoint-based topology optimization of unsteady fluid flows using the lattice {Boltzmann} method},
	volume = {4},
	doi = {10.1299/mej.17-00120},
	number = {3},
	journal = {Mechanical Engineering Journal},
	author = {Chen, Cong and Yaji, Kentaro and Yamada, Takayuki and Izui, Kazuhiro and Nishiwaki, Shinji},
	year = {2017},
	note = {Num Pages: 17-00120},
	pages = {17--00120},
}

@article{ito_geometry_2026,
	title = {Geometry reconstruction from magnetic resonance velocimetry measurements via solving an inverse fluid flow problem},
	issn = {0021-9991},
	url = {https://www.sciencedirect.com/science/article/pii/S0021999126005036},
	doi = {10.1016/j.jcp.2026.115151},
	urldate = {2026-06-24},
	journal = {Journal of Computational Physics},
	author = {Ito, Shota and Zimmermann, Alexander and Jeßberger, Julius and Simonis, Stephan and Kummerländer, Adrian and Bukreev, Fedor and Thöming, Jorg and Pesch, Georg and Krause, Mathias J.},
	month = jun,
	year = {2026},
	pages = {115151},
}

@article{ito_identification_2024,
	title = {Identification of reaction rate parameters from uncertain spatially distributed concentration data using gradient-based {PDE} constrained optimization},
	volume = {167},
	issn = {0898-1221},
	url = {https://www.sciencedirect.com/science/article/pii/S0898122124002451},
	doi = {10.1016/j.camwa.2024.05.026},
	urldate = {2026-06-24},
	journal = {Computers \& Mathematics with Applications},
	author = {Ito, Shota and Jeßberger, Julius and Simonis, Stephan and Bukreev, Fedor and Kummerländer, Adrian and Zimmermann, Alexander and Thäter, Gudrun and Pesch, Georg R. and Thöming, Jorg and Krause, Mathias J.},
	month = aug,
	year = {2024},
	pages = {249--263},
}

@misc{ito_generation_2026,
	address = {Rochester, NY},
	type = {{SSRN} {Scholarly} {Paper}},
	title = {Generation of efficient adjoint lattice {Boltzmann} methods with algorithmic differentiation},
	url = {https://papers.ssrn.com/abstract=6505987},
	doi = {10.2139/ssrn.6505987},
	language = {en},
	urldate = {2026-06-24},
	publisher = {Social Science Research Network},
	author = {Ito, Shota and Kummerländer, Adrian and Jeßberger, Julius and Grafen, Johannes L. and Öz, Ecem and Gauger, Nicolas R. and Sagebaum, Max and Krause, Mathias J.},
	month = apr,
	year = {2026},
}

@article{laniewski-wollk_adjoint_2016,
	title = {Adjoint {Lattice} {Boltzmann} for topology optimization on multi-{GPU} architecture},
	volume = {71},
	issn = {0898-1221},
	url = {https://www.sciencedirect.com/science/article/pii/S0898122115006215},
	doi = {10.1016/j.camwa.2015.12.043},
	number = {3},
	urldate = {2026-06-24},
	journal = {Computers \& Mathematics with Applications},
	author = {{Łaniewski-Wołłk, Ł. and Rokicki, J.}},
	month = feb,
	year = {2016},
	pages = {833--848},
}

@article{dugast_topology_2018,
	title = {Topology optimization of thermal fluid flows with an adjoint {Lattice} {Boltzmann} {Method}},
	volume = {365},
	issn = {0021-9991},
	url = {https://www.sciencedirect.com/science/article/pii/S0021999118302067},
	doi = {10.1016/j.jcp.2018.03.040},
	urldate = {2026-06-24},
	journal = {Journal of Computational Physics},
	author = {Dugast, Florian and Favennec, Yann and Josset, Christophe and Fan, Yilin and Luo, Lingai},
	month = jul,
	year = {2018},
	pages = {376--404},
}

@article{krause_adjoint-based_2013,
	series = {Mesoscopic {Methods} in {Engineering} and {Science}},
	title = {Adjoint-based fluid flow control and optimisation with lattice {Boltzmann} methods},
	volume = {65},
	issn = {0898-1221},
	url = {https://www.sciencedirect.com/science/article/pii/S0898122112005421},
	doi = {10.1016/j.camwa.2012.08.007},
	number = {6},
	urldate = {2026-06-24},
	journal = {Computers \& Mathematics with Applications},
	author = {Krause, Mathias J. and Thäter, Gudrun and Heuveline, Vincent},
	month = mar,
	year = {2013},
	pages = {945--960},
}

@article{pingen_adjoint_2009,
	title = {Adjoint parameter sensitivity analysis for the hydrodynamic lattice {Boltzmann} method with applications to design optimization},
	volume = {38},
	issn = {0045-7930},
	url = {https://www.sciencedirect.com/science/article/pii/S0045793008001989},
	doi = {10.1016/j.compfluid.2008.10.002},
	number = {4},
	urldate = {2026-06-24},
	journal = {Computers \& Fluids},
	author = {Pingen, Georg and Evgrafov, Anton and Maute, Kurt},
	month = apr,
	year = {2009},
	pages = {910--923},
}

@article{klemens_solving_2020,
	series = {Mesoscopic {Methods} in {Engineering} and {Science}},
	title = {Solving fluid flow domain identification problems with adjoint lattice {Boltzmann} methods},
	volume = {79},
	issn = {0898-1221},
	url = {https://www.sciencedirect.com/science/article/pii/S0898122118303754},
	doi = {10.1016/j.camwa.2018.07.010},
	number = {1},
	urldate = {2026-06-24},
	journal = {Computers \& Mathematics with Applications},
	author = {Klemens, Fabian and Förster, Benjamin and Dorn, Márcio and Thäter, Gudrun and Krause, Mathias J.},
	month = jan,
	year = {2020},
	pages = {17--33},
}

@article{liu_discrete_2014,
	title = {Discrete adjoint sensitivity analysis for fluid flow topology optimization based on the generalized lattice {Boltzmann} method},
	volume = {68},
	issn = {0898-1221},
	url = {https://www.sciencedirect.com/science/article/pii/S0898122114004507},
	doi = {10.1016/j.camwa.2014.09.002},
	number = {10},
	urldate = {2026-06-24},
	journal = {Computers \& Mathematics with Applications},
	author = {Liu, Geng and Geier, Martin and Liu, Zhenyu and Krafczyk, Manfred and Chen, Tao},
	month = nov,
	year = {2014},
	pages = {1374--1392},
}

@article{yaji_topology_2016,
	title = {Topology optimization in thermal-fluid flow using the lattice {Boltzmann} method},
	volume = {307},
	issn = {0021-9991},
	url = {https://www.sciencedirect.com/science/article/pii/S0021999115008244},
	doi = {10.1016/j.jcp.2015.12.008},
	urldate = {2026-06-24},
	journal = {Journal of Computational Physics},
	author = {Yaji, Kentaro and Yamada, Takayuki and Yoshino, Masato and Matsumoto, Toshiro and Izui, Kazuhiro and Nishiwaki, Shinji},
	month = feb,
	year = {2016},
	pages = {355--377},
}

@article{krause_openlbopen_2021,
	series = {Development and {Application} of {Open}-source {Software} for {Problems} with {Numerical} {PDEs}},
	title = {{OpenLB}—{Open} source lattice {Boltzmann} code},
	volume = {81},
	issn = {0898-1221},
	url = {https://www.sciencedirect.com/science/article/pii/S0898122120301875},
	doi = {10.1016/j.camwa.2020.04.033},
	urldate = {2026-06-24},
	journal = {Computers \& Mathematics with Applications},
	author = {Krause, Mathias J. and Kummerländer, Adrian and Avis, Samuel J. and Kusumaatmaja, Halim and Dapelo, Davide and Klemens, Fabian and Gaedtke, Maximilian and Hafen, Nicolas and Mink, Albert and Trunk, Robin and Marquardt, Jan E. and Maier, Marie-Luise and Haussmann, Marc and Simonis, Stephan},
	month = jan,
	year = {2021},
	pages = {258--288},
}

@article{kummerlander_efficient_vocal_2026,
	title = {Efficient fluid structure interaction simulation of vocal fold oscillations using a homogenized {Lattice} {Boltzmann} {Method}},
	volume = {457},
	issn = {0045-7825},
	url = {https://www.sciencedirect.com/science/article/pii/S0045782526002823},
	doi = {10.1016/j.cma.2026.119009},
	urldate = {2026-06-24},
	journal = {Computer Methods in Applied Mechanics and Engineering},
	author = {Kummerländer, Adrian and Tur, Bogac and Haase, Maik and Bukreev, Fedor and Döllinger, Michael and Krause, Mathias J. and Kniesburges, Stefan},
	month = aug,
	year = {2026},
	pages = {119009},
}

@article{teutscher_digital_2025,
	title = {A digital urban twin enabling interactive pollution predictions and enhanced planning},
	volume = {281},
	issn = {0360-1323},
	url = {https://www.sciencedirect.com/science/article/pii/S0360132325005748},
	doi = {10.1016/j.buildenv.2025.113093},
	urldate = {2026-06-24},
	journal = {Building and Environment},
	author = {Teutscher, Dennis and Bukreev, Fedor and Kummerländer, Adrian and Simonis, Stephan and Bächler, Peter and Rezaee, Ashkan and Hermansdorfer, Mariusz and Krause, Mathias J.},
	month = aug,
	year = {2025},
	pages = {113093},
}

@article{kummerlander_efficient_2026-1,
	title = {Efficient wall-modelled large eddy simulation of rotors using homogenized lattice {Boltzmann} methods},
	volume = {36},
	issn = {0961-5539},
	url = {https://doi.org/10.1108/HFF-09-2025-0724},
	doi = {10.1108/HFF-09-2025-0724},
	number = {7},
	urldate = {2026-06-24},
	journal = {International Journal of Numerical Methods for Heat \& Fluid Flow},
	author = {Kummerländer, Adrian and Ito, Shota and Schecher, Maximilian and Dapelo, Davide and Simonis, Stephan and Krause, Mathias J. and Bukreev, Fedor},
	month = apr,
	year = {2026},
	pages = {2649--2673},
}

\end{document}